\documentclass[fleqn,usenatbib]{mnras}

\usepackage[T1]{fontenc}
\usepackage{xcolor}
\DeclareRobustCommand{\VAN}[3]{#2}
\let\VANthebibliography\thebibliography
\def\thebibliography{\DeclareRobustCommand{\VAN}[3]{##3}\VANthebibliography}

\usepackage{graphicx}	
\usepackage{amsmath}	
\usepackage{amssymb}	
\usepackage{newtxtext,newtxmath}

\title[A major merger caught by eROSITA]{A major galaxy cluster merger caught by eROSITA: weak lensing mass distribution and kinematic description}

\author[Rog\'erio Monteiro-Oliveira]{
Rog\'erio Monteiro-Oliveira$^{1}$\thanks{E-mail: rogerionline@gmail.com}
\\
$^{1}$Universidade Estadual de Santa Cruz, Depto. de Ci\^encias Exatas e Tecnol\'ogicas, Lab. de Astrof\'isica Te\'orica e Observacional, 45650-000, Ilh\'eus, BA, Brazil\\ 
}

\date{Accepted XXX. Received YYY; in original form ZZZ}

\pubyear{2015}

\begin{document}
\label{firstpage}
\pagerange{\pageref{firstpage}--\pageref{lastpage}}
\maketitle

\begin{abstract}

We present the weak lensing mass distribution of a triple merging cluster candidate at $z_{\rm photo}\sim 0.36$ belonging to a supercluster recently discovered during the eROSITA Performance Verification phase. Our analysis solved a previous tension in the merger classification by confirming that the cluster pair eFEDS J093513.3+004746 and eFEDS J093510.7+004910 is undergoing a major merger with a mass ratio $1.7_{-0.7}^{+0.5}$. According to our two-body kinematic description, the encounter happened  $0.58_{-0.20}^{+0.15}$ Gyr ago,  in a scenario that supports the observed radio relic position at the cluster outskirts. However, the same analysis showed that the companion cluster, eFEDS J093501.1+005418, is not gravitationally bound to the interacting system and therefore it is not part of the supercluster. We also checked the impact of adopting a scaling relation to determine the halo concentration $c_{200}$. At the observed merger stage, where the clusters have travelled $\sim$55 per cent of the path to reach the apoapsis, the choice of the $c_{200}$ (whether from a scaling relation or a free parameter in the mass model) does not change significantly either the cluster masses or the kinematic description.

\end{abstract}




\section{Introduction}
\label{sec:intro}

Given the large amount of energy involved \citep[$\gtrsim 10^{64}$ ergs; e.g.,][]{sarazin04}, major galaxy cluster mergers\footnote{We adopt the definition of \cite{martel14}, and consider as major merger an event in which the mass ratio between the two most massive (in case of multiple clusters) is less than two.} are often called astrophysical particle colliders \citep{harvey15}. Therefore, they are an excellent laboratory to investigate  properties of the key constituents of clusters, namely the intracluster medium gas \citep[ICM; e.g.][]{Keshet21}, the galaxies \citep[e.g enhancement/quenching of star formation;][]{Kelkar20, Hernandez-Lang22}, and the dark matter \citep[e.g.][]{Fischer21}, even if they are observed at late stages of a merger \citep[e.g.,][]{Tam20b}. As the ICM  carries the most discernible signatures of the merger process \citep[such as shock waves, cold fronts, and sloshing; e.g.,][]{Ha18,Pandge19,Ueda20,Ueda21,Cho21,Machado22}, ongoing high-quality X-ray surveys as the extended ROentgen Survey with an Imaging Telescope Array \citep[eROSITA;][]{Merloni12, eROSITA} will dramatically increase the number of identified cluster mergers.

Analysing the eROSITA Final Equatorial Depth Survey (eFEDS), \citet[][hereafter G21]{Ghirardini21}, reported the discovery of a supercluster at photometric redshift (photo-$z$) $z_{\rm photo}\sim 0.36$. It consists of a chain of eight clusters spanning a length of 27 Mpc on the plane of the sky. Despite the fact that most of them present regular X-ray and optical properties, the authors identified a triple major merger candidate consisting the galaxy clusters eFEDS J093513.3+004746 (hereafter eFEDS4746; $z_{\rm photo}=0.367$), eFEDS J093510.7+004910 (hereafter eFEDS4910; $z_{\rm photo}=0.367$) and eFEDS J093501.1+005418 (hereafter eFEDS5418; $z_{\rm photo}=0.385$). More details of these clusters are presented in Table~\ref{tab:masses}. Using LOFAR and uGMRT observations, G21 found two radio relics at the outskirts of eFEDS4746 and eFEDS4910 as well as a radio halo coincident with those clusters in projection. Radio relics are characterized by diffuse radio synchrotron emission, which are known to be good tracers of the shock wave propagating though the ICM after the pericentric passage during a merger \citep[][]{Zhang20,Knowles22}, even though their  origin are not yet fully understood \citep[e.g.][]{vanWeeren19}. On the other hand, due to its regular optical and X-ray properties, together with the absence of any radio feature, G21 argued that eFEDS5418 is in a pre-merger state given its short projected distance to its companions.

In contrast with the radio observations, the major merger scenario between eFEDS4746 and eFEDS4910 does not find any support either from the estimated clusters' richness or mass ratios, $\sim$4:1. It is also worth  mentioning that mass estimates based on the $L_X-M$ scaling relations, like those provided by G21, can be highly biased for clusters involved in a merger \citep[e.g.][]{Torri04}. Unfortunately, G21 stated that the shallow data from  eROSITA prevented them to extract more details about the merger dynamics, because the X-ray counterparts of the radio relics cannot be detected. To solve this discrepancy in the merger classification, an essential piece of information is the map of the underlying dark matter distribution from which a comprehensive description of the current merger status can be obtained.

\begin{table*}
\caption[]{Summary of the cluster properties found by G21. The original $M_{500}$ were converted to $M_{200}$ assuming the halo mass density profile is well described by an NFW profile with a concentration parameter $c_{200}$ given by the $M_{200}$--$c_{200}$ scaling relation presented by \cite{duffy08}.}
\label{tab:masses}
\begin{center}
\begin{tabular}{l c c c c}
\hline
\hline 
 Full name & Short name & $z_{\rm photo}$  & Richness & $M_{200}$ \\
           &            &      &  & ($10^{14}$ M$_\odot$)\\
\hline
eFEDS J093501.1+005418 & eFEDS5418 & 0.385 & $143\pm12$ & $4.2\pm0.8$  \\ [5pt]
eFEDS J093510.7+004910 & eFEDS4910 & 0.367 & $62\pm9$ & $2.2\pm1.1$ \\ [5pt]
eFEDS J093513.3+004746 & eFEDS4746 & 0.367 & $208\pm15$ & $8.9\pm1.7$ [ \\ 
\hline
\hline
\end{tabular}
\end{center}
\end{table*}

In this work, we use the public data release of the Hyper Suprime-Cam Subaru Strategic Program (HSC-SSP; \citealt{HSCSSP.PDR2}) to perform the first total mass reconstruction in the field containing these clusters based on the weak gravitational lensing technique. Next, we measure the masses of individual  clusters and check if the major merger scenario is feasible. We also address the kinematics of the merger through an analytical two-body description and examine whether the positions of the radio relics are consistent with the proposed merger history.   Recently, \citet[][hereafter C22]{Chadayammuri21}   pointed that dark matter halo shapes can dramatically change during the merger event,  therefore the use of a scaling relation to determining the halo concentration, $c_{200}$ \citep[e.g.,][]{duffy08},  in the mass modeling could lead to an overestimate of cluster mass. To investigate the possible impact on the mass determinations and the kinematic description, we test two scenarios, where in one we include  $c_{200}$ as a free parameter in the model, while in the other a scaling relation is assumed.

The paper is organised as follows. In Section~\ref{sec:mass} we describe the weak lensing analysis and the respective results. The proposed scenario for the merger kinematic is detailed in Section~\ref{sec:dyn.kin}. Then, our findings are discussed in Section~\ref{sec:discussion} and sumarized in Section~\ref{sec:summary}. In this paper, we adopt the standard $\Lambda$CDM cosmology, with parameters $\Omega_m = 0.27$, $\Omega_\Lambda = 0.73$, $\Omega_k = 0$, and $h=0.7$.

\begin{figure*}
\begin{center}
	\includegraphics[width=1.0\textwidth]{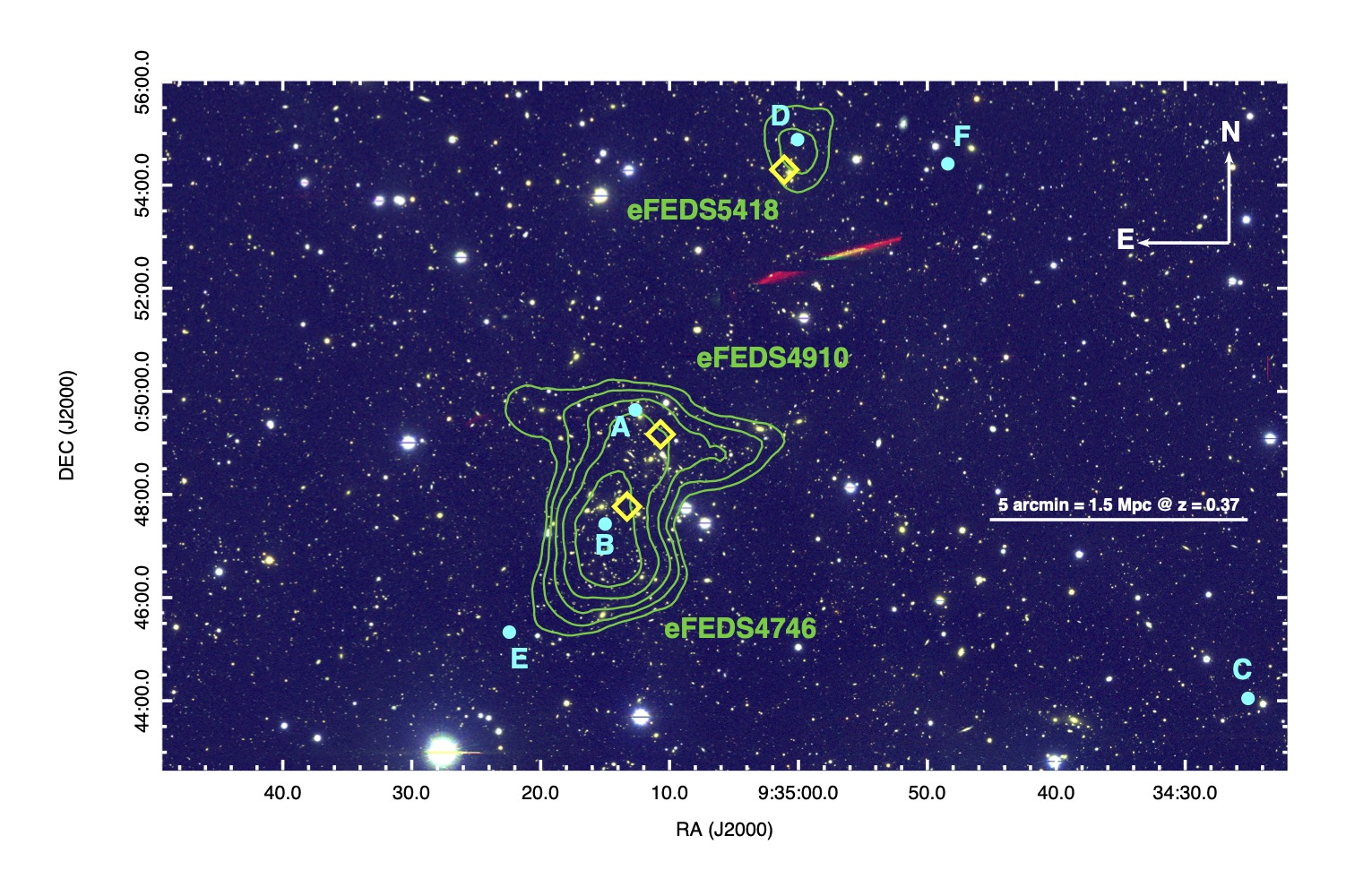}
    \caption{Composite $gri$ image showing the location of the three galaxy clusters investigated in this work (labelled in the figure). This is only part of a larger field for which the data (imaging and photometric catalogue) were extracted from the HSC-SSP PDR2. The yellow diamonds are placed at the galaxy clusters' positions as identified by G21, based on photo-$z$s. The green contours show the projected density distribution of the red sequence galaxies selected through a statistical subtraction method in the colour-colour space (more details given in Section~\ref{sec:gal.pop}). The pair eFEDS4746 and eFEDS4910 forms a bimodal system, possibly in interaction, and, together with eFEDS5418 will have their kinematic investigated in this work in order to confirm (or not) if a triple merger is taking place. The cyan dots labeled A--F represent the positions of mass peaks identified in the weak lensing analysis presented in this work (Section~\ref{sec:mass.map}). Each of the above-mentioned clusters coincides with a mass peak (A, B,  and D, respectively).}
    \label{fig:field}
\end{center}
\end{figure*}

\section{Projected mass reconstruction}
\label{sec:mass}

\subsection{Data}
\label{sec:data}

The data underlying this work was retrieved from the wide layer of the second public data release of the HSC-SSP\footnote{Available at \url{https://hsc-release.mtk.nao.ac.jp/doc/index.php/sample-page/pdr2/}}. The survey is now completed, covering $\approx 1100$ deg$^2$ of the sky at $z<1.5$, with five broad bands $grizy$ reaching a depth of 5$\sigma$ for a point source at $i_{\rm AB} \approx 26$.

To accomplish our goals of measuring the galaxy shapes with the highest precision possible, we downloaded the $i$-band imaging from the Data Archive System (DAS\footnote{\url{https://hsc-release.mtk.nao.ac.jp/das\_search/pdr2/}}). We focused on a region with $0.5\times0.5$ deg$^2$ in area centred at RA = 09:35:12, DEC = +00:48:00, roughly matching the position of eFEDS4746. The optical photometric catalogue ($griz$) and the underlying photo-z's \citep[$z_{\rm photo}\equiv$~{\sc photo-z\_best}; ][]{Tanaka18,Nishizawa20} were also retrieved for all objects in the field.

\subsection{Red sequence selection}
\label{sec:gal.pop}

Despite accounting for only $\approx$5 per cent of the total mass, the stellar content of galaxies works as a good qualitative tracer of the cluster's gravitational potential, which is dominated by the dark matter. The projected density of galaxies allows for the characterization of the cluster's morphology (e.g., if the system is uni- or multi-modal) and, ultimately, the verification of the correspondence between the baryonic and dark matter, through a comparison of the projected density of galaxies and that of dark matter, the latter obtained via gravitational lensing \citep[e.g.,][]{Monteiro-Oliveira17b, Monteiro-Oliveira20, Wittman17}.

\begin{figure*}
\begin{center}
	\includegraphics[width=\textwidth]{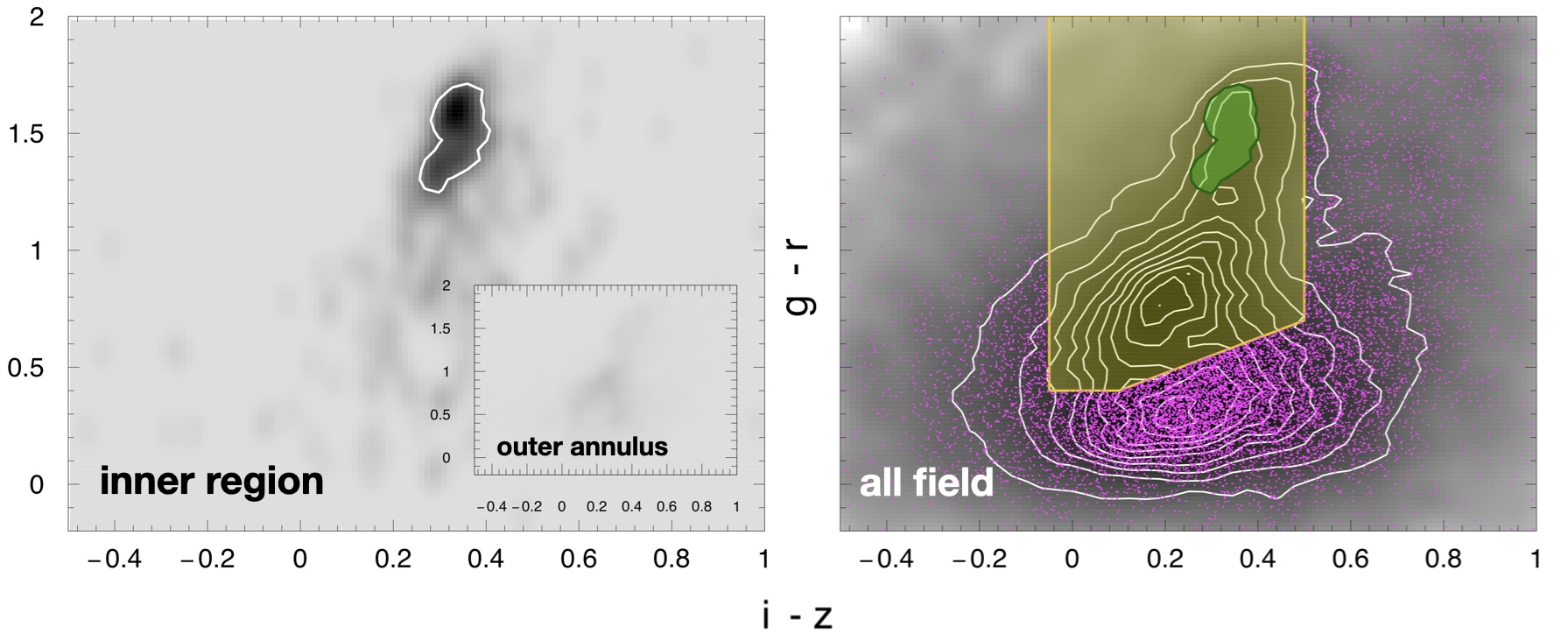}
    \caption{{\it Left: } $i-z$ vs $g-r$ colour-colour (CC) diagram of  galaxies lying in two different locations of the field. The inner region, where red sequence galaxies are expected to dominate in terms of number, 
    was selected as a circle with radius of $2'$ centered at eFEDS4746.
    We also selected  galaxies in an outer annulus, $10' \leq {\rm radius} \leq 14'$ from eFEDS4746, shown in the inset. After a statistical subtraction of the CC diagrams, we defined the cluster locus (white contour) preferably inhabited by red sequence member galaxies (Section~\ref{sec:gal.pop}). {\it Right: } CC diagram of all galaxies in the field. The cluster locus is highlighted in green. For the weak lensing analysis, we are only interested in the galaxies located behind the cluster  ($z> z_{\rm cluster}$), and any contribution from both the red sequence and foreground galaxies ($z< z_{\rm cluster}$) will decrease the lensing signal-to-noise (S/N). We defined an empirical foreground locus (yellow polygon) and considered as potential source candidates all galaxies located outside.  The magenta points represent the final source sample selected after  some quality cuts in the shape  parameters (Section~\ref{sec:source}).}
    \label{fig:cc}
\end{center}
\end{figure*}

\begin{figure}
\begin{center}
	\includegraphics[width=\columnwidth]{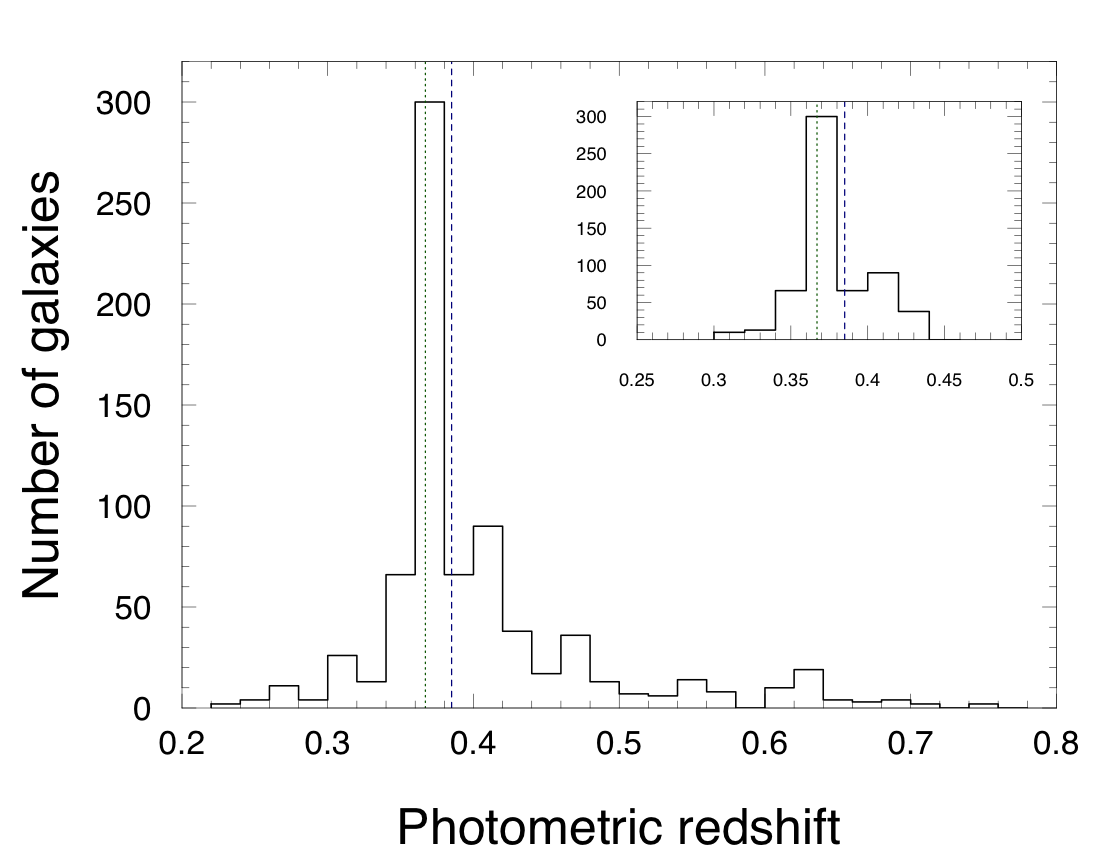}
    \caption{Photometric redshift distribution of the 765 galaxies identified as red sequence members through the statistical subtraction in the CC space.  The vertical dotted (dashed) line shows the location of eFEDS4748/eFEDS4910 (eFEDS5418) as found by G21. The inset panel corresponds to those galaxies matching the selection criteria of \protect\citet[][i.e., $0.31\leq z_{\rm photo} \leq 0.44$]{wen13}  considering the lowest and highest redshift limits of the three clusters. This sub-sample comprises 76 per cent of the CC-selected red sequence members.}
    \label{fig:cluster.photo-z}
\end{center}
\end{figure}

The selection of  galaxies belonging to the clusters' red sequence \citep{Visvanathan77} was done through the application of the statistical subtraction technique in the $g-r$ versus $i-z$ colour-colour (CC) space. Two heterogeneous regions were considered in this process: the first one centred at eFEDS4748 (a circular region with a radius of $2'$) in which red cluster member galaxies are supposed to be numerically dominant, and a distant, ``control'' annulus ($10'  \leq {\rm radius} \leq 14' $), where the counts of the field galaxies are expected to be dominant.  As red sequence galaxies have similar photometric properties at a given redshift \citep[e.g.,][]{Stott09}, we expect them to form a well-defined locus in the CC plane, as  can be seen in Fig.~\ref{fig:cc}. The locus boundaries become even more prominent after subtracting the contribution of the outer region, allowing for a confident CC selection.  In the end, we found  765 red sequence members.  Their photo-$z$ distribution, presented in Fig.~\ref{fig:cluster.photo-z}, shows a prominent peak around the clusters' location, suggesting the selection was reasonably accurate. For the sake of comparison, we applied the photo-$z$ cut introduced by \citet{wen13}\footnote{According to these authors, cluster members correspond to all galaxies inside the slice $z_{\rm cluster} - 0.04(1+z_{\rm cluster}) \leq z \leq z_{\rm cluster} + 0.04(1+z_{\rm cluster})$}, finding that 3/4 of our red sequence members match this criteria.

We built the projected red sequence distribution after smoothing the discrete field with the function
\begin{equation}
    D(\vec{\xi}) = \sum_{i=1}^N  K(\vec{\xi_i},\sigma_\xi)\ \mbox{,}
\label{eq:dens}
\end{equation}
which represents a sum over all $N$ galaxies with radial coordinate $\vec{\xi}$ located inside the smoothing scale $\sigma_\xi$, set to $1'$. We adopted the Epanechnikov kernel,
\begin{align}
K(\vec{\xi_{i}},\sigma_\xi) = \left \{
\begin{array}{ll}
\frac{3}{4} \left [1-\left(\frac{\vec{\xi_{i}}}{\sigma_\xi}\right)^2 \right ], & \vec{\xi_{i}} \le \sigma_\xi \\
0, & \vec{\xi_{i}} > \sigma_\xi.
\end{array}
\right.
\label{eq:kernel}
\end{align}
The resulting galaxy surface density map, shown in Fig.~\ref{fig:field}, is fully consistent with the cluster positions given by G21. It clearly shows a  bimodal distribution, corresponding to the pair eFEDS4748 and eFEDS4910, and a third clump reasonable away from the previous overdensity that is related to eFEDS5418. An interesting feature is the two prickle-like structures emanating from eFEDS4910, which could be the end of large scale filaments that we believe to be feeding galaxy clusters \citep{Kuchner22}.

\subsection{Shape measurements}
\label{sec:shape}

The weak lensing effect is described in terms of two quantities, the convergence
\begin{equation}
\kappa=\frac{\Sigma(\rm{\xi})}{\Sigma_{\rm cr}}\ \mbox{,}
\label{eq:kappa}
\end{equation}
and the shear,
\begin{equation}
\gamma= \gamma_1+i\gamma_2\ \mbox{.}
\label{eq:shear}
\end{equation}

The former, a scalar, reflects the projected mass density of the gravitational lens and describes the change  in size on background galaxies (note that the surface brightness is conserved at the same time). It is written in units of the lensing critical density,
\begin{equation}
\Sigma_{\rm cr}=\frac{c^2 D_{\rm s} }{4\pi G D_{\rm ds} D_{\rm d}}\ \mbox{,}
\label{eq:sigma.cr}
\end{equation}
where $D_{\rm s}$, $D_{\rm ds}$ and $D_{\rm d}$ are, respectively, the angular diameter distances to the source\footnote{Another way to refer to background galaxies}, between the lens and the source, and to the lens. 

The second quantity, a spin-2 tensor, refers to the image stretching. Similar to the convergence, the shear is related to the projected gravitational potential of the lens \citep[e.g.,][]{Umetsu20}. The total effect caused by the lens is called the reduced shear, and involves a combination of convergence and shear,
\begin{equation}
g \equiv \frac{\gamma} {1-\kappa}  \ \mbox{.} 
\label{eq:red.shear}
\end{equation}
The role of the galaxy cluster is to induce a coherent distortion on the source galaxies,  changing their  ellipticity in the sense that, on averaged,
\begin{equation}
\langle e \rangle \simeq g  \ \mbox{.} 
\label{eq:wl.regime}
\end{equation}
By definition, weak gravitational lensing is a statistical phenomenon which means it can only be measured over a large sample of background galaxies.

As the shape parameters of the galaxies in our region of interest are not made publicly available at the time this work was been carried out, we measured them by ourselves. We started by checking the image quality. Despite the source galaxies being the only objects carrying the gravitational lensing signal, the identification of non-saturated stars are crucial for the evaluation imaging quality and the assessment of the point spread function (PSF).

We built our own photometric catalogue of the $i$-band imaging by running the software {\sc SExtractor} \citep{sextractor} for object identification. Then, the star/galaxy classification was done on their full width half-maximum (FWHM) as the following:  objects within $0.60''\leq {\rm FWHM} \leq 0.71''$  were classified as stars (point sources) whereas those having ${\rm FWHM} > 0.73''$  were considered as galaxies (extended objects). Our final galaxy catalogue comprises all objects matched with those from the HSC-SSP catalogue having the highest confident photo-$z$ estimations \citep[{\sc photo-z\_conf\_best}~$>0.13$; ][]{Medezinski18}

\begin{figure}
\begin{center}
	\includegraphics[width=\columnwidth]{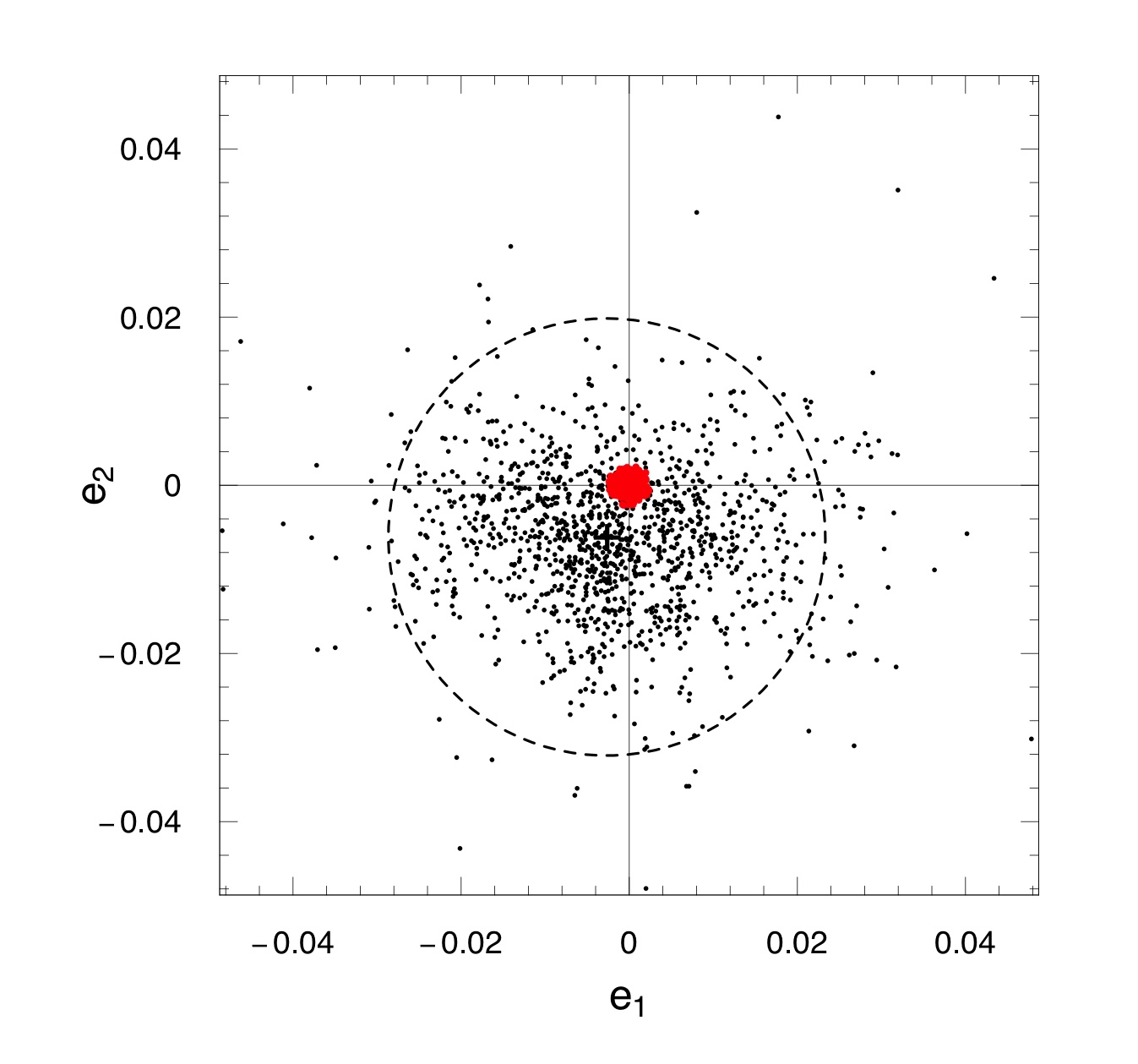}
    \caption{PSF modelling based on bright and unsaturated stars homogeneously distributed across the field. The 1024 black dots show the raw distribution of the ellipticity components $e_1$ and $e_2$, with $\langle e_1\rangle=-0.002\pm0.013$ and $\langle e_2\rangle=-0.006\pm0.009$. After removing outliers to create an analytical function to describe the PSF across the image, we found the residuals represented by the 934 red points as low as $0\pm7\times10^{-4}$ for both $e_1$ and $e_2$.} 
    \label{fig:psf}
\end{center}
\end{figure}

The Bayesian code {\sc im2shape} \citep{bridle98} was adopted to measure the ellipticity components $e_1$ and $e_2$ of stars and galaxies. It works by modelling the objects as a sum (or a single, in the case of stars) of Gaussian functions. Given that the observed unsaturated star profiles are the result of the convolution between the PSF with a  delta function, the discrete set of ($e_1$, $e_2$) can be spatially interpolated across the image to create an analytical description of the PSF. We accomplished this task resorting to the {\sc Thin Place Regression} \citep[][]{fields} function in the {\sc R}  environment \citep{R}. The interpolation was done three times, each one removing the 10 per cent worst objects (i.e., those with the largest absolute residuals). At the end, we arrived at a tight fit as manifested by the very small residuals ($0.0000\pm 0.0007 $ for $e_1$ and $e_2$) as shown in Fig.~\ref{fig:psf}.

The galaxy ellipticities were then extracted after  {\sc im2shape} has performed the PSF deconvolution. For the sake of quality, we disregarded all galaxies having measured ellipticity   uncertainties greater than 0.2, or showing any evidence of blending. Unfortunately, as we are not able to know a priori the shapes of the unlensed galaxies, $e_1$ and $e_2$ are at best noisy proxies of the shear field. When required, we adopted $\sigma_{\rm int}=0.3$ as the dispersion of the intrinsic ellipticity distribution of source galaxies \citep[e.g.][]{Leauthaud07}.

\subsection{Source selection}
\label{sec:source}

The cluster and the foreground members have to be carefully removed from the source sample, in order not to critically dilute the weak lensing signal. To this end, we tested three selection criteria, one based on a CC cut, and two others relying on the galaxies photo-$z$, but excluding the cluster locus in one of them. Using the examples presented in \cite{med10} as guidelines, we defined an empirical region in the CC diagram comprising the most probable location of foreground candidates plus the cluster locus (Section~\ref{sec:gal.pop}). The CC based selection process, illustrated in Fig.~\ref{fig:cc}, resulted in a source density of 13.2 galaxies per arcmin$^{-2}$. For the photo-$z$ based selection, we considered all galaxies beyond the upper limit suggested by \citet[][i.e., $z_{\rm photo}> 0.44$]{wen13}, yielding 16.1 galaxies per arcmin$^{-2}$. Excluding galaxies within the cluster locus, the density decreases to 15.6 galaxies per arcmin$^{-2}$.

We referred to the mass aperture statistic \citep{schneider96} to create a shear signal-to-noise map,  
\begin{equation}
{\rm S/N}=\dfrac{\sqrt{2}}{\sigma_{\rm int}^2}\ \dfrac{\sum_{i=1}^{N_{\theta_0}} e_{+_i}(\theta_i) Q_{\rm NFW}(\theta_i,\theta_0)}{\left[ \sum_{i=1}^{N_{\theta_0}} Q_{\rm NFW}^2(\theta_i,\theta_0)\right ]^{1/2}} \ \mbox{,}
\label{eq:SN}
\end{equation} 
based on the tangential ellipticity, 
\begin{equation}
\begin{aligned}
& e_+  =-e_1\cos(2\phi)-e_2\sin(2\phi) \mbox{,} \\
& \phi = \arctan \left (  \frac{y_i-y_{\rm bin}}{x_i-x_{\rm bin}} \right)\ \mbox{,}
\end{aligned}
\label{eq:et}
\end{equation} 
averaged over the $N_{\theta_0}$ galaxies inside a circular region of $\theta_0=8'$  computed in each of the $1.5\times10^2$ spatial bins centred at the Cartesian coordinates $x_{\rm bin}, y_{\rm bin}$. The other quantities in Eq.~\ref{eq:SN} are the radial position of the $i$-th galaxy $\theta_i$ and the dispersion of the intrinsic ellipticity distribution $\sigma_{\rm int}$. We adopted a filter \citep{Schirmer04} that roughly matches an NFW shear profile, 
\begin{multline}
\noindent Q_{\rm NFW}(\theta_i,\theta_0)=[1+e^{a-b\chi(\theta_i,\theta_0)}+e^{-c+d\chi(\theta_i,\theta_0)}]^{-1} \times  \\
\dfrac{\tanh [\chi(\theta_i,\theta_0)/\chi_c]}{\pi\theta_0^2[\chi(\theta_i,\theta_0)/\chi_c]}\mbox{,}
\label{eq:nfw.filter}
\end{multline}
where  $\chi=\theta_i/\theta_0$. Following \cite{hetterscheidt05}, we adopted $a=6$, $b=150$, $c=47$, $d=50$, and $\chi_c=0.15$.

\begin{figure*}
\begin{center}
	\includegraphics[width=\textwidth]{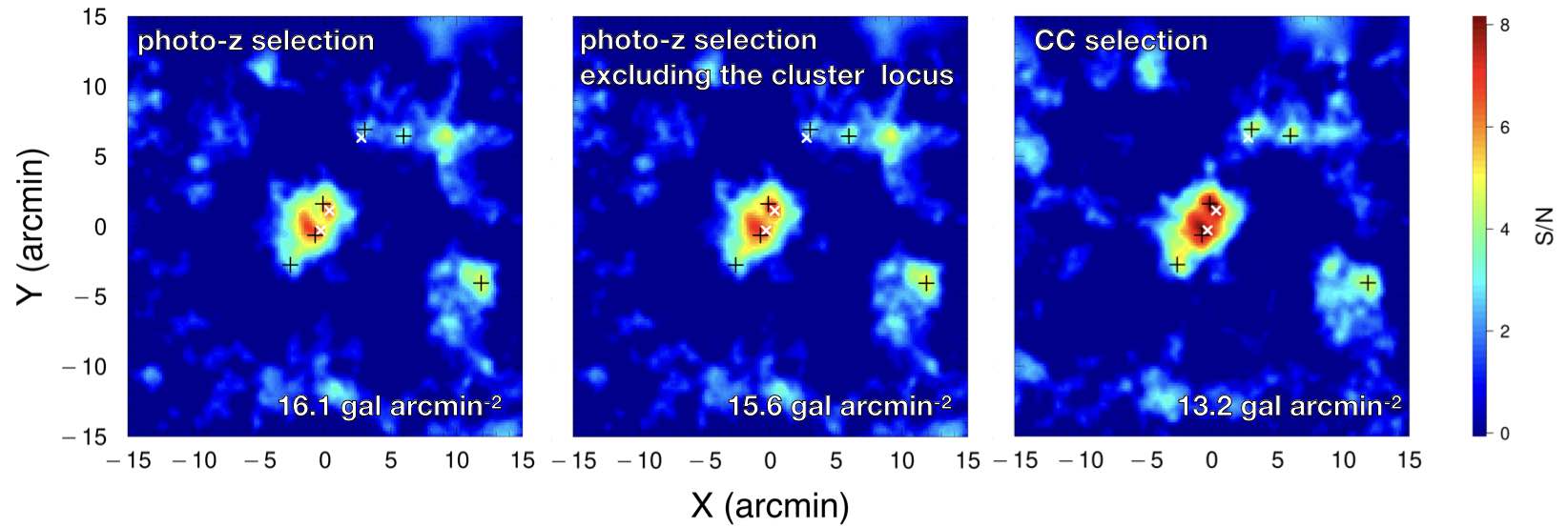}
    \caption{Weak lensing mass aperture statistic S/N. As guidelines, the white $\times$ signs show the positions of the clusters as found by G21. The black $+$ signs mark the positions of the most relevant mass peaks according to the analysis presented in Section~\ref{sec:mass.map}. We tested three different methods for selecting the source galaxies; the first is solely based on photo-$z$ ($z_{\rm photo}> 0.44$, left), the second being a hybrid one, where we excluded from the previous sample all galaxies inside the cluster locus (centre), and the third one, which is only based on the CC cuts (Fig.~\ref{fig:cc}, right). The resulting source density is indicated in the lower right corner. We showed that the CC-based selection maximises the S/N in the vicinity of the interacting cluster candidates eFEDS4748 and eFEDS4910 (see details in the text), and is thus regarded as our fiducial sample.}
    \label{fig:sn}
\end{center}
\end{figure*}

The resultant S/N maps based on the three selection methods are presented in Fig.~\ref{fig:sn}. 
Overall the three maps appear very similar, and
all of them have the highest S/N in the same region of the clusters eFEDS4746 and eFEDS4910. When comparing the maps quantitatively within a circular region with $4'$ radius  centred at eFEDS4746, we found that the CC based map reaches the highest S/N among the samples. The 97.5 per cent percentile (maximum) of the S/N is 7.65 (8.10), 6.61 (7.17) and 6.24 (6.92) respectively for the CC based, the photo-$z$ minus the cluster locus and the photo-$z$ samples. This conclusion is in line with \cite{Medezinski18}, who found that CC cuts are more efficient than photo-$z$'s in removing  contributions from both cluster and foreground galaxies and thus maximizing  the weak lensing signal. We will adopt the CC-based selection as our fiducial source sample henceforth. The 36,589 source galaxies correspond to a critical density  $\Sigma_{\rm cr}=2.8\pm 0.6 \times 10^9$ M$_\odot$ kpc$^{-2}$ (Eq.~\ref{eq:sigma.cr}). The quoted error on the critical density reflects the spread of the source redshifts, which is characterized by a mean photo-$z$ of $\approx 1.1$.

\subsection{Projected mass field and modelling}
\label{sec:mass.map}

The Bayesian code {\rm LensEnt2} \citep{LensEnt2} was applied to translate the source ellipticities into the cluster projected mass distribution. In a nutshell, this maximum entropy algorithm works by maximizing the evidence of the reconstructed mass field with respect to the data. Since each individual galaxy is a noisy proxy of the shear and is correlated with its neighbourhood, an intrinsic correlation function (ICF) must be applied to smooth the data. We adopted a Gaussian ICF \citep[e.g.][]{Monteiro-Oliveira18,Monteiro-Oliveira20,Monteiro-Oliveira21} with a FWHM $\sigma_{\rm ICF}=90''$. The final convergence map is presented in Fig.~\ref{fig:field}. The noise level, $\sigma_\kappa=0.04$, was calculated as follows. For each of 100 iterations, the mass map was computed after every galaxy orientation was rotated by a random angle in the interval [0,180]  to remove the lensing signal.

The convergence map is fully consistent with the mass aperture statistic S/N (Fig.~\ref{fig:sn}). We identified six significant mass concentrations (i.e.,  $\geq4\sigma_\kappa$), labelled A--F. The clumps A, B and D can be correlated respectively to the galaxy clusters eFEDS4910 (RA = 09:35:13,  DEC = +00:49:38), eFEDS4746 (RA = 09:35:15 , DEC = +00:47:26) and eFEDS5418 (RA = 09:35:00, DEC = +00:54:53) as they show a good agreement with the cluster locations reported by G21 (diamonds in Fig~\ref{fig:field} and crosses in Fig.~\ref{fig:sn}) and also match the distribution of cluster red sequence members (Fig.~\ref{fig:field}). Given the lack of any optical counterpart in the field, the peaks C, E,  and F probably consist of a combination of structures seen in projection through the line-of-sight \citep[e.g.,][]{Liu16,Wei18}.

The next step was to measure the individual halos masses. The gravitational lensing signal induced in each source galaxy corresponds to the sum of the effect due to $N$ halos, $\kappa = \sum_{i=1}^{N}\kappa_i$, $\gamma_j = \sum_{i=1}^{N}\gamma_{i}$, with  $j={1,2}$. We assumed that the halos density profile can be described by the NFW profile \citep{nfw96}. In this case, the model encompasses four basic parameters, namely the lens centre ($x_c,y_c$), halo mass $M_{200}$\footnote{The total mass enclosed within a sphere whose density is 200 times the critical density of the Universe,  $\rho_c=\frac{3H^2(z)}{8\pi G}$.} and  concentration $c_{200}$.

When modelling the halo masses, we were also interested in checking if
the way halos are selected affects the final mass measurements. For example, is it possible that a model that only considers halos with a secure optical counterpart would produce discrepant mass measurements than a model that accounts for all high S/N halos? To address this question, three models were computed. In model \#1, we measured all halos masses (assuming $z_{\rm photo}\approx 0.36$ for all), whereas in model \#2, only the three known clusters had their masses computed. In both cases, we set the halo concentration using the $M_{200}-c_{200}$ scaling relation proposed by \cite{duffy08}. Aiming to check the potential impact of the merger age on the measured concentration, in model \#3, we included $c_{200}$ as a free parameter when measuring the halo masses. The full parameter vectors $\Theta$ are presented in Table~\ref{tab:model.desc}. For all models, the halo centres were kept fixed at the position found by our algorithm.

\begin{table}
\begin{center}
\caption{Models description. $N_{\rm par}$ corresponds to the number of parameters in each model. $M$ and $c$ refer respectively to $M_{200}$ and $c_{200}$.}
\begin{tabular}{lccc}
\hline
\hline
Model & Halos & $\Theta$ & $N_{\rm par}$\\
\hline  
\#1 & 6 (A--F) & $M^{\rm A}$, $M^{\rm B}$, $M^{\rm C}$, $M^{\rm D}$, $M^{\rm E}$, $M^{\rm F}$ & 6\\
\#2 & 3 (A, B, D) & $M^{\rm A}$, $M^{\rm B}$, $M^{\rm D}$ & 3 \\
\#3 & 3 (A, B, D) & $M^A$, $M^B$, $M^D$, $c^{\rm A}$, $c^{\rm B}$, $c^{ \rm D}$ & 6\\
\hline
\end{tabular}
\label{tab:model.desc}
\end{center}
\end{table}

The $\chi^2$ statistic for each model is
\begin{equation}
\chi^2=\sum_{j=1}^{N_{{\rm sources}}} \sum_{i=1}^{2}  \frac{[g_i(M_{200},x_c,y_c)-e_{i,j}]^ 2}{\sigma_{\rm int}^2+\sigma_{{\rm obs}_{i,j}}^2}, 
\label{eq:chi2.shear}
\end{equation}
where $g_i$ is the theoretical reduced shear (Eq.~\ref{eq:red.shear}), $e_{i,j}$ is the measured ellipticity of source galaxies (Sec.~\ref{sec:shape}), and $\sigma_{{\rm obs}_{i,j}}$ is the error on shape measurement given by {\sc im2shape}. The likelihood is
\begin{equation}
\mathcal{L} \propto \exp \left (-\frac{\chi^2}{2}\right)  \, .
\label{eq:Log.lik}
\end{equation}
Finally, we write the posterior of our problem as 
\begin{equation}
\noindent {\rm Pr}(\Theta|{\rm data}) \propto \mathcal{L}({\rm data}|\Theta) \times \mathcal{P}(\Theta)  \, .
\label{eq:posterior}
\end{equation}
For the models \#1 and \#2, we applied a flat prior $\mathcal{P}(\Theta)$ for the masses, $0<M_{200}\leq 10^{16}$ M$_\odot$, to avoid non-physical values and to accelerate the convergence. In model \#3 we also added a prior on the concentration, $0<c_{200}\leq 15$.

The posterior in Equation~\ref{eq:posterior} was sampled by the MCMC algorithm with a Metropolis sampler 
MCMCMETROP1R \citep[][]{MCMCpack}. Four chains of $10^5$ elements plus $10^4$ as 'burn-in' were generated for each model, all of which were considered convergent at the end \citep{coda}. The marginalized posteriors are shown in Table~\ref{tab:model}.

\begin{table*}
\caption[]{Results of the modelling. The median was taken as the representative value for each posterior, and the quoted error corresponds to the 68 per cent c.l. interval. Masses $M_{200}$ are presented in units of $10^{14}$\,M$_\odot$. The concentrations shown in parenthesis are those predicted by \protect\cite{duffy08}. Based on AIC and BIC statistics, model \#1 was considered as the preferred one.}
\label{tab:model}
\begin{center}
\begin{tabular}{l c c c c c c c c c c c c}
\hline
\hline 
Model & $M_{200}^{\rm A}$ & $M_{200}^{\rm B}$ & $M_{200}^{\rm C}$ & $M_{200}^{\rm D}$ & $M_{200}^{\rm E}$ & $M_{200}^{\rm F}$ & 
$c_{200}^{\rm A}$ & $c_{200}^{\rm B}$ & $c_{200}^{\rm D}$ &  $\Delta$AIC & $\Delta$BIC\\
 & eFEDS4910 & eFEDS4746 &  & eFEDS5418 &  &  & 
 &  &  &  &  & \\
\hline

$   \#1$	 & 
                               $2.63_{-1.18}^{+0.96}$	 & 
                               $2.73_{-1.48}^{+1.08}$	 &
                               $4.23_{-1.81}^{+1.53}$	 &
                               $2.43_{-1.30}^{+1.07}$	 &
                               $1.27_{-0.94}^{+0.61}$	 &
                               $3.08_{-1.53}^{+1.20}$	 &
                               $(3.4_{-0.1}^{+0.1})$   &
                               $(3.4_{-0.2}^{+0.1})$ &
                               $(3.4_{-0.2}^{+0.1})$ &
                               0 &
                               2    \\[5pt]
                               
$   \#2$	 & 
                               $2.54_{-1.23}^{+0.96}$	 & 
                               $3.46_{-1.52}^{+1.28}$	 &
                               -- &
                               $3.42_{-1.55}^{+1.19}$	 &
                               -- &
                               -- &
                               $(3.4_{-0.1}^{+0.1})$   &
                               $(3.3_{-0.1}^{+0.1})$ &
                               $(3.3_{-0.1}^{+0.1})$ &
                               20 &
                               0    \\[5pt]
$   \#3$	 & 
                               $2.80_{-1.70}^{+1.21}$	 & 
                               $3.27_{-1.95}^{+1.44}$	 &
                               -- &
                               $2.32_{-1.14}^{+0.87}$	 &
                               -- &
                               -- &
                               $2.9_{-1.7}^{+1.2}$	 &
                               $2.6_{-1.4}^{+1.0}$	 &
                               $7.1_{-3.9}^{+3.0}$	 &
                               23 &
                               26    \\
\hline
\hline
\end{tabular}
\end{center}
\end{table*}

It is remarkable that, regardless of the model, the masses of all eFEDS clusters are comparable within the error bars. To select the best model, we resorted to the Akaike Information Criterion (AIC\footnote{${\rm AIC}=2k-2\ln{\mathcal{\Hat{L}}}$, where $k$ is the number of parameters in the considered model and $\mathcal{\Hat{L}}$ is the corresponding maximum log-likelihood }) and the Bayesian Information Criterion (BIC\footnote{${\rm BIC}=k\ln{n}-2\ln{\mathcal{\Hat{L}}}$, where $n$ is the number of data points.}). Both metrics are based on the idea that simplest models are preferred over the most complex ones, penalising therefore those with a large number of parameters. A small difference between the two is that BIC also takes into account the number of data points, thus avoiding overfitting. Among a finite number of models, those with the lowest AIC/BIC is considered the best  to describe the data.

The model \#1 is the preferred according to the AIC \citep[$\Delta{\rm AIC}>10$;][]{kass95} whereas the BIC statistic suggests that it is comparable to model \#2 ($\Delta{\rm BIC}<2$). These results make us confident in choosing the model \#1 as the fiducial one henceforth. The full posterior plot, presented in Fig.~\ref{fig:triangle}, shows that, in general, the parameters are not (anti-)correlated among them, except when the nearest neighbour(s) is considered (A -- B, B -- E, D -- F). Nevertheless, even in these cases, the anti-correlation is very weak. Another remarkable feature in the model is the considerably high mass associated with an isolated halo, labelled ``C'', surpassing even the identified cluster masses. Similarly, for clumps ``E'' and ``F'', we did not find any clear optical counterparts after computing the photo-$z$ distribution of galaxies inside a circular region with a radius of $1.5'$ ($\sim$ 0.5 Mpc), implying that they probably do not correspond to a single halo. \cite{Wei18} showed that a collection of halos more massive than $10^{13}$ M$_\odot$ ($z \approx 1$) can be detected at line-of-sight with a ${\rm S/N} \geq 3$. We end this discussion with an important caveat: the masses in Table~\ref{tab:model} only correspond to the ``true'' value if the halo redshift is the same as that considered in the modeling, $z=0.36$.

\begin{figure}
\begin{center}
	\includegraphics[width=\columnwidth]{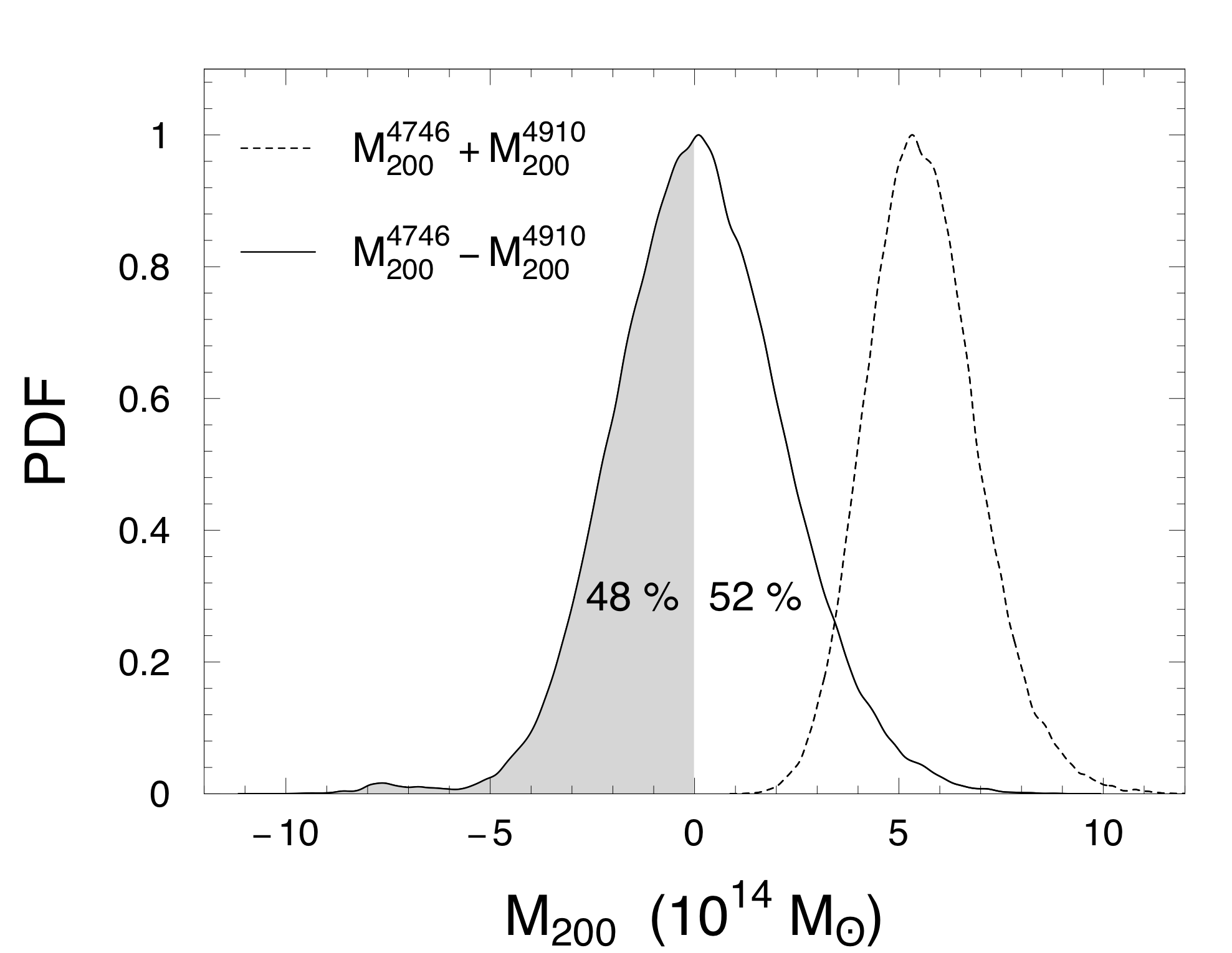}
    \caption{Probability density function (PDF) of the sum (dashed line) and the difference (M$_{200}^{4746}$ -- M$_{200}^{4910}$, continuous line)  of the posteriors of masses of individual clusters. In 52 (48) per cent of the $4\times10^5$ MCMC samples, eFEDS4748 (eFEDS4910) emerges as the most massive cluster of the pair. The total mass is $5.53_{-1.37}^{+1.24}\times10^{14}$ M$_\odot$.}
    \label{fig:sum}
\end{center}
\end{figure}

\begin{figure*}
\begin{center}
	\includegraphics[width=\textwidth]{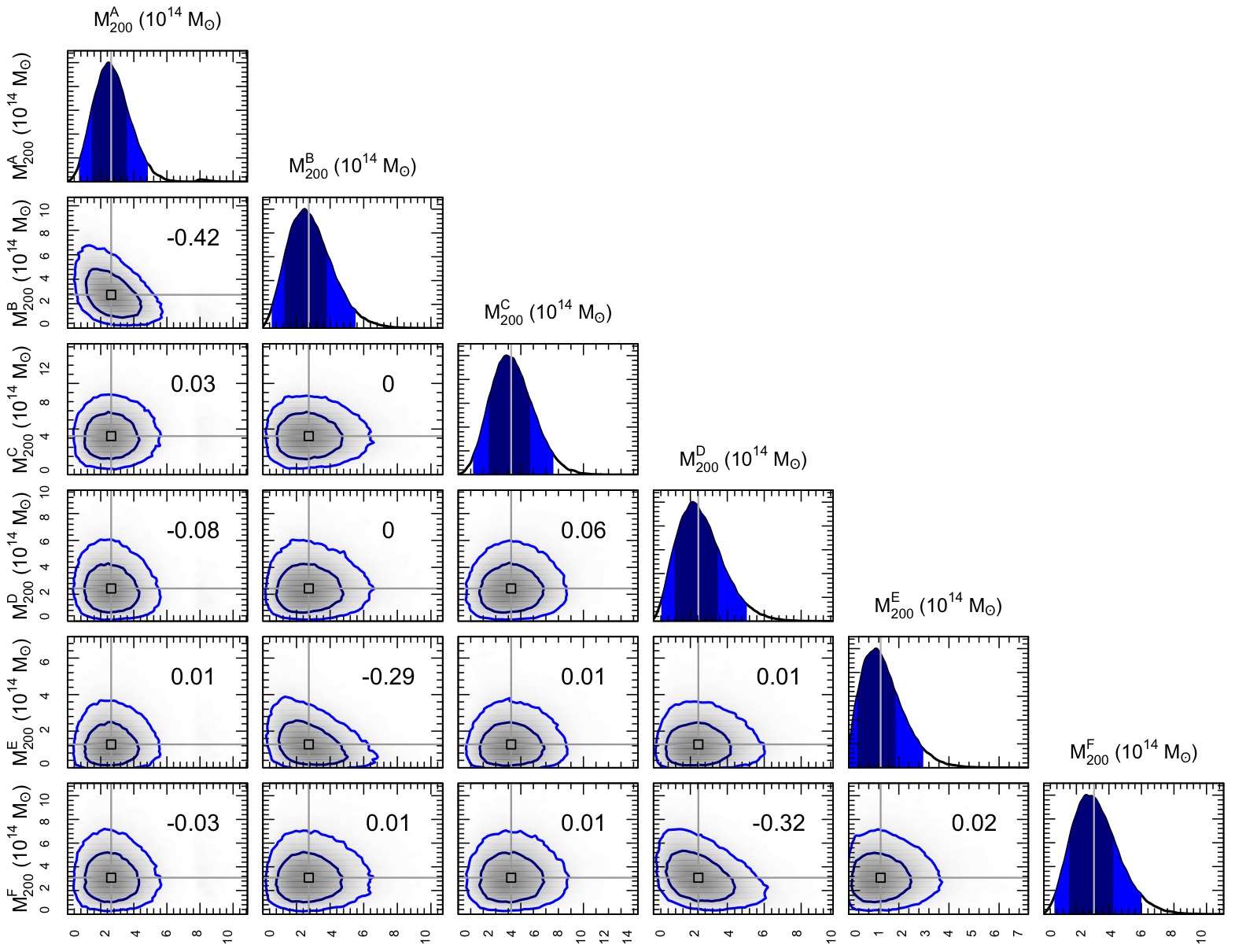}
    \caption{Posteriors of the parameters of model \#1  (six NFW-halos) as mapped by the MCMC sampler. Dark (light) blue corresponds to 68 (95) per cent c.l.  Along the diagonal line  the marginalised posteriors of each halo (A -- F) are shown. We considered the median as the representative value for each distribution. The number inside the plots shows the Pearson's correlation coefficient for each pair.}
    \label{fig:triangle}
\end{center}
\end{figure*}

Regarding the interacting binary system candidate, eFEDS4746 is the more massive one in 52\% of the MCMC samples, as we can see in Fig.~\ref{fig:sum}. Regardless of which cluster is considered the most massive, the mass ratio (i.e. the ratio between the corresponding posteriors, $\mathcal{R} \equiv {\rm M}_{200, \rm cluster}/ {\rm M}_{200, \rm subcluster}$) is $\mathcal{R} = 1.7_{-0.7}^{+0.5}$.  In 63 per cent of the samples, we found $\mathcal{R}<2$, suggesting that the system can be classified as a major merger, while in another 28 per cent, $2\leq \mathcal{R}<4$, suggesting that a semi-major merger class is more suitable to describe the possible interaction \citep{martel14}. If we consider a toy model in which eFEDS4746 and eFEDS4910 will merge within a few Gyr \citep[e.g.][]{Machado+2015}, the final cluster would have a mass of $M_{200}=5.53_{-1.37}^{+1.24}\times10^{14}$ M$_\odot$ (Fig.~\ref{fig:sum}), considering the progenitors masses are conserved.

The halo concentrations are fully consistent in both models \#1 and \#2. When setting $c_{200}$ as a free parameter in  model \#3, we found comparable values within the large error bars, despite the face values being systematically smaller than those obtained by the scaling relation.  The exception is eFEDS5418, with a considerably higher concentration.

\section{Kinematic analysis}
\label{sec:dyn.kin}

\begin{figure}
\begin{center}
	\includegraphics[width=\columnwidth]{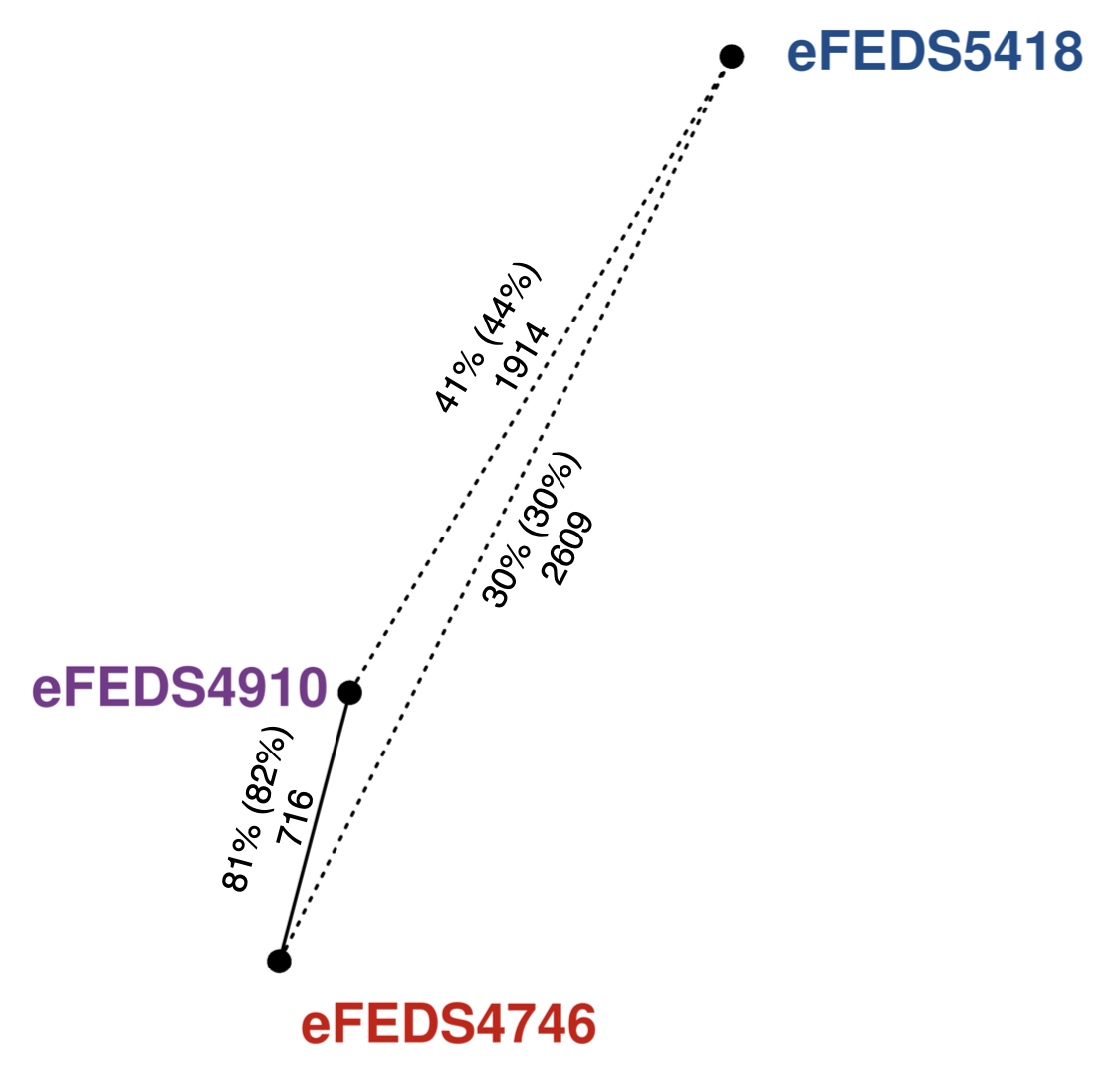}
    \caption{A sketch showing the likely configuration of the galaxy clusters studied in this work. The dots correspond to their respective halo centre position according to our weak lensing analysis. The percentages refer to the probability of each pair being found gravitationally bound according to the MCMAC code. The same quantity, but computed with the masses delivered by the model \#3 (i.e. $c_{200}$ is given by the modelling), is shown in parenthesis.  When the drawn line is continuous (dashed) the pair is considered bound (unbound). The number below refers to the projected separation between the mass peaks (in kpc).}
    \label{fig:skeleton}
\end{center}
\end{figure}

Tailored hydrodynamical simulations have been proven to provide a realistic description of the merger kinematics  \citep[e.g.,][]{Molnar20,Doubrawa20, Moura21}. However, when it is not possible to utilize such a tool, we have to resort to a simplified yet still reliable analytical description of the two-body interaction, such as the Monte Carlo Merger Analysis Code \cite[MCMAC;][]{dawson}. This is a Bayesian model designed to calculate the possible merger scenarios from the probability density function (PDF) with just a few input parameters, namely the cluster masses ($M_{200}$, Table~\ref{tab:model}), their spatial separation projected at an angle  $\alpha$ from the plane of the sky ($R_{\rm p}$, Fig~\ref{fig:skeleton}) and their mean redshift ($\bar{z}$, Table~\ref{tab:masses}). The great improvements offered by  this code over other tools are the treatment of the clusters as spatially extended objects instead of point masses \citep{beers82} and the availability of the posteriors of the quantities of interest (e.g., time since/to the pericentric passage, maximum separation). MCMAC assumes mass conservation (i.e., $M=M_1+M_2={\rm constant}$) during the zero impact parameter merger and no angular momentum. The maximum relative velocity $V_{\rm r}$ is set to be the free-fall velocity according to the cluster masses. It has two versions, one designed for post mergers (MCMAC-post) and another for pre-interaction pairs (MCMAC-pre). In the latter case, the condition for bound pairs,
\begin{equation}
V_{\rm r}^2R_{\rm p}\leq 2GM\sin^2 \alpha \cos \alpha,
\label{eq:condition}
\end{equation}
is not necessarily satisfied during the MCMC re-samplings. The probability of a system being bound (unbound) can be defined as the ratio between the number of MCMC-pre samples in which  Eq.~\ref{eq:condition} is true (false) and the total number of samples, 10,000 in this work (for both MCMAC-pre/post). We adopted the cluster redshifts estimated in G21. To compute the uncertainties, we assumed a squared sum of a typical error on subcluster redshift \citep[$\sigma_z\approx 0.007$; e.g.,][]{Monteiro-Oliveira17a,Monteiro-Oliveira18} with an uncertainty of 5 per cent corresponding to the photo-$z$ precision of the HSC-SSP data \citep{Tanaka18}, which will have a much larger weight on the final estimation. We also considered a realistic prior on the velocity of the merger component on the plane of sky \citep[$v_{\rm plane}<1500$ km s$^{-1}$;][]{Monteiro-Oliveira22}.

Despite the fact that MCMAC is designed to describe two-body interactions, in \cite{Monteiro-Oliveira22} we suggested that the algorithm can be applied in more complex systems, by reducing them to one pair at a time. Although not dealing with all bodies in the system simultaneously, this approach provides a realist chronological order for the collision events. In this sense, we use the MCMAC-pre mode to describe the potential mergers, as illustrated in Fig.~\ref{fig:skeleton}.

We found that only the pair eFEDS4746/4910 is gravitationally bound, a scenario recovered in 81 per cent of the MCMC-pre samples. However, when either of these clusters forms a pair with eFEDS5418, a bound system is only reported by less than 41 per cent of the samples. 
Even when we increased the mass of the cluster to the value of the combined mass of eFEDS4746 and eFEDS4910 in our toy model, the correspondingly increased percentage (48\%) is not enough to change the kinematic classification. Regarding any possible impact on the mass reconstruction described with a free $c_{200}$ parameter (model \#3 in Table~\ref{tab:model}), all of the results presented in this paragraph remain consistent.

Our proposed post-merger kinematic description is presented in Table~\ref{tab:post}.  MCMAC-post does not distinguish between two possible scenarios: (1) a collision that has just happened $TSC_0$ Gyr ago and the members are outgoing, or (2) a system that already reached the apoapsis and the clusters are incoming for a new encounter $TSC_1$ Gyr after the last pericentric passage. This degeneracy can be broken, however, if any other age proxy is also observed. The presence of radio relics in the outskirts of eFEDS4746/4910 is itself a piece of strong evidence that the clusters have had their encounter not too long ago \citep[e.g., A3376;][]{Machado13, Monteiro-Oliveira17b}. Therefore, based on this observable, we disregarded the incoming scenario in which the collision happened $\sim$3.13 Gyr ago. Thus, our kinematic description suggests that the collision between eFEDS4746 and eFEDS4910 took place $0.58_{-0.20}^{+0.15}$ Gyr ago with a 3-dimensional velocity of $2148_{-402}^{+393}$ km s$^{-1}$ in a direction orientated $42_{-23}^{+24}$ degrees from the plane of the sky. The system is been observed after running $\sim$55 per cent of the path to the apoapsis, where the clusters will be $1.8_{-1.1}^{+0.9}$ Mpc apart. Again, these results remain almost unchanged when considering the alternative model \#3 with the halo  concentration as a free parameter.

\begin{table}
\caption[]{Kinematic description of the post-merger scenario for the cluster pair eFEDS4748--eFEDS4910 according to  MCMAC-post. The first five rows show the input quantities (from top to down): the cluster masses $M$, the cluster redshifts $z$, and their projected separation $d_{\rm proj}$. The following 7 rows are:  the parameter estimations corresponding to the angle between the merger axis and the plane of the sky, $\alpha$, the current 3D relative velocity, $v_{\rm 3D,obs}$, the current 3D separation between the clusters, $d_{\rm 3D,obs}$, the 3D velocity at collision time, $v_{\rm 3D,col}$, the 3D maximum separation at the clusters' apoapsis, $d_{\rm 3D,max}$, the time since collision for the outgoing scenario, $TSC_0$, and the time since collision for the incoming scenario,  $TSC_1$.}
\label{tab:post}
\begin{center}
\begin{tabular}{l c | c c }
\hline
\hline
Quantity & Unit &  Median	&	68 per cent c.l \\
\hline
$M_{200}^{\rm 4746}$	  &  	$10^{14}$ M$_\odot$	  &  	2.96	 & 	1.55 -- 4.32	  \\ [5pt]
$M_{200}^{\rm 4910}$	  &  	$10^{14}$ M$_\odot$	  &  	2.74	 & 	1.58 -- 3.85	  \\ [5pt]
$z^{\rm 4746}$	  &  	--	  &  	0.367	 & 	0.351 -- 0.382	  \\ [5pt]
$z^{\rm 4910}$	  &  	--	  &  	0.367	 & 	0.352 -- 0.382	  \\ [5pt]
$d_{\rm proj}$	  &  	Mpc	  &  	0.71	 & 	0.66 -- 0.75	  \\ [5pt]
$\alpha$	  &  	degrees	  &  	42	 & 	19 -- 66	  \\ [5pt]
$v_{\rm 3D,obs}$	  &  	km s$^{-1}$	  &  	1194	 & 	724 -- 1667	  \\ [5pt]
$d_{\rm 3D,obs}$	  &  	Mpc	  &  	0.96	 & 	0.65 -- 1.18	  \\ [5pt]
$v_{\rm 3D,col}$	  &  	km s$^{-1}$	  &  	2148	 & 	1747 -- 2541	  \\ [5pt]
$d_{\rm 3D,max}$	  &  	Mpc	  &  	1.8	 & 	0.7 -- 2.7	  \\ [5pt]
$TSC_0$	  &  	Gyr	  &  	0.58	 & 	0.38 -- 0.72	  \\ [5pt]
$TSC_1$	  &  	Gyr	  &  	3.13	 & 	0.87 -- 5.81	  \\ 
\hline
\hline
\end{tabular}
\end{center}
\end{table}

\section{Discussion}
\label{sec:discussion}

\begin{figure*}
\begin{center}
	\includegraphics[width=0.9\textwidth]{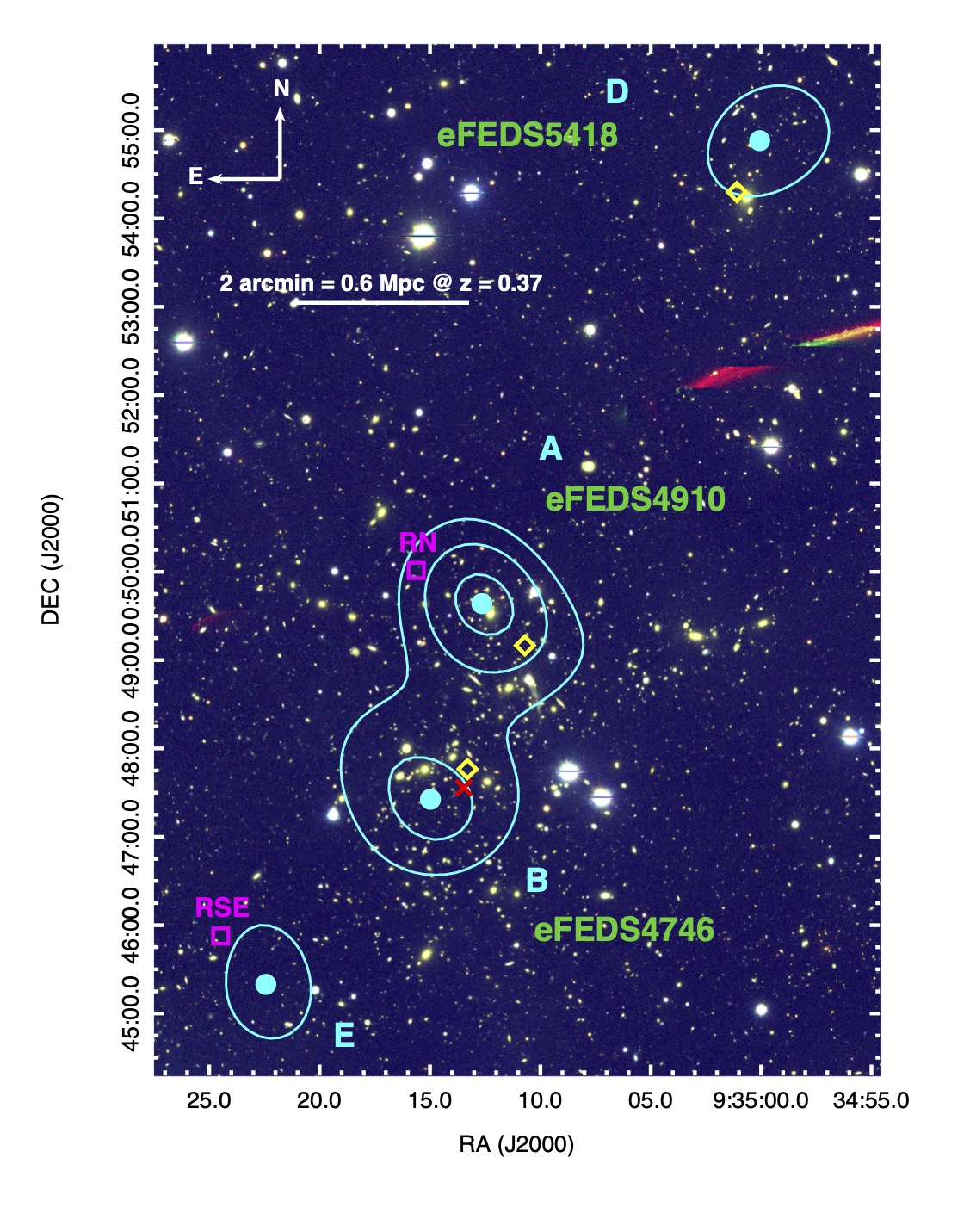}
    \caption{Summary of the analysis presented in this work. The $gri$ image shows a zoomed-in view of the three clusters. The yellow diamonds are cluster coordinates according to G21, and the red X is the position of the single X-ray peak. Our weak lensing analysis has identified six significant mass peaks  (i.e., $\geq4\sigma_\kappa$, A--F in decreasing order of significance), whose locations are shown by the cyan points (halos C and F are outside of the current field-of-view). Three of them correspond to the known clusters, eFEDS4910 (A), eFEDS4746 (B), and eFEDS5418 (D). At the outskirts of eFEDS4746/4910, there are two radio relics (RSE and RN; magenta squares), a hint of a recent merger activity. The cyan contours represent the convergence map in units of $\sigma_\kappa$, starting from $4\sigma_\kappa$.}
    \label{fig:discussion}
\end{center}
\end{figure*}

Our weak lensing mass map shows an excellent agreement with the cluster positions detected by eROSITA and Subaru (Fig.~\ref{fig:discussion}). Assuming that the  density profile of each cluster can be  satisfactorily described by an NFW profile, we found that all the clusters have similar masses (Table~\ref{tab:model}). For the last two in the Table, these estimates show a good agreement with G21, considering the error bars. The most outstanding feature arising from our weak lensing analysis is the absence of a dominant cluster, contrary to the  findings of G21. Their mass estimate for eFEDS4746 based on the $L_X-M$ scaling relation is $\sim3$ times larger than our  value. Even though their estimates  agree with those from the {\it Planck} catalogue, we stress that the weak lensing technique does not make any assumption a priori about the cluster dynamical state \citep[e.g.,][]{Soja18,Umetsu20}.

The ICM distribution is asymmetric in the interacting pair, showing a single peak related to eFEDS4746 ('X' in Fig.~\ref{fig:discussion}), located $\sim23''$  away from the corresponding mass peak. As this separation is less than the expected error on the mass peak centroid caused by the effects of shape noise and smoothing of the mass map \citep{dietrich12}, we can not make any assumption about any possible spatial coincidence (or not) between the mass and gas peaks. On the other hand, the lack of an X-ray peak leads us to conclude that eFEDS4910 has temporarily lost its gas counterpart, a feature only seen in the class of dissociative mergers \citep{dawson}. Even so, such a highly asymmetric gas post-merger configuration is supported by the observed mass ratio,  $1.7_{-0.7}^{+0.5}$, as the ICM dynamics is mostly governed by the initial conditions of the gas distribution \citep[e.g. concentration;][]{Machado+2015,Moura21}.

Since there are no spectroscopic redshift measurements for these galaxy clusters, we have to resort to their photometric redshifts estimated by G21 in order to address their kinematics.  This was accomplished with an analytical two-body analysis applied to each pair to infer  the possible pre- and post-merger scenarios. We confirmed that the pair eFEDS4746/4910 is highly likely bounded.  On the other hand, the  cluster eFEDS5418 is not gravitationally bound to either of eFEDS4746/4910 (when considered  individually).  This conclusion remains the same even in the case when we consider a fictional cluster formed from the merger between eFEDS4746/4910 (having the combined masses) placed midway their current positions, although the probability of the unbound scenario decreases to 52 per cent. Therefore, our analysis does not support the conjecture proposed by G21 that eFEDS5418 is a pre-merger and part of the supercluster. Regarding the last feature, this is not surprising since the selection criterion adopted by G21 is based solely on the local over-density instead of a more physically motivated one \citep[e.g., occurrence of a future gravitational collapse;][]{chon15b}.

Our proposed kinematic description posits that eFEDS4746 and eFEDS4910  are undergoing a major merger orientated with an axis of $42_{-23}^{+24}$ degrees from the plane of the sky. Despite the error bars, such a large angle can explain the absence of any brightness and temperature jumps at the positions of both radio relics, as these features can be detected even at angles as large as 30 degrees (C22). The pericentric passage happened $0.58_{-0.20}^{+0.15}$ Gyr ago. We can also provide an alternative estimation of the merger age, by computing the time required for the shock to travel from the cluster centre  (in this case the midpoint with RA = 09:35:14, DEC = +00:49:39) to the radio relic positions. Taking the relic RNE, whose projected location ($\sim 1.2$ Mpc) matches the expectation for a cluster merger \citep[$\sim 1$ Mpc; e.g.,][]{Ha18} and considering a shock velocity of $v\sim3000$ km s$^{-1}$ \citep{Machado+2015}, we arrive at $\sim0.5$ Gyr, in agreement with the kinematic estimation. Considering the relic RN instead ($\sim0.5$ Mpc), we find a considerable lower age, $\sim0.2$ Gyr. Despite both being consistent with a merger caught not so far from the pericentric passage, the differences in distance to the cluster centre of the radio relics could be a hint that they were not generated during the same event.

The halo concentration $c_{200}$ can significantly change during the several phases of the cluster merger, mainly in the moments close to the pericentric passage (C22). Therefore, it could have an important role in the cluster mass measurement (C22) and consequently on the whole merger kinematic description, depending on whether it is determined from a scaling relation \citep[obtained from relaxed halos;][]{duffy08} or included as a free parameter in the modeling. After testing the two scenarios, we found comparable masses and therefore, similar merger histories. A possible explanation for such an agreement is, at the current merger phase (i.e., at the current time since the pericentric passage), any changes in $c_{200}$ have completely vanished. Unfortunately, a direct comparison with the temporal evolution of $c_{200}$ presented in C22 (see their Fig.~4) is not possible, as the authors presented it in terms of an arbitrary time, starting from the beginning of the simulation.

\section{Summary}
\label{sec:summary}

Based on the second public data release of the Hyper Suprime-Cam Subaru Strategic Program (HSC-SSP PDR2), we present the first weak gravitational lensing reconstruction of the  matter distribution in the Northern part of the eFEDS supercluster, where \cite{Ghirardini21} found an interacting cluster pair (eFEDS4746/4910) surrounded by a companion (eFEDS5418). The masses of the clusters are found to be $2.73_{-1.48}^{+1.08}\times10^{14}$ M$_\odot$ (eFEDS J093513.3+004746), $2.63_{-1.18}^{+0.96}\times10^{14}$ M$_\odot$ (eFEDS J093510.7+004910), and $2.43_{-1.30}^{+1.07}\times10^{14}$ (eFEDS J093501.1+005418), based on our analysis. Our result supports the scenario drawn from the X-ray and radio observations, in which eFEDS4746 amd eFEDS4910 are undergoing a  major merger. However, eFEDS5418 does not show any evidence of being gravitationally bound to the pair. We also found no impact of the halo concentration on the mass modeling and kinematic description, 
irrespective of the method employed (i.e., from a scaling relation or being a free parameter of the lensing model).
A spectroscopic follow-up of the cluster members can offer a complementary view of the merger dynamics and corroborate our results, which are based solely on  photometric redshifts.

\section*{Acknowledgements}

I want to thank an anonymous referee for his/her valuable comments that helped improve this paper. My thanks also go to Prof. Yen-Ting Lin for revising this manuscript and to ASIAA, where this work was completed,  for the hospitality.

This work was conceived and developed during the most acute phase of the COVID-19 pandemic. Although many people on the front line had to be exposed daily to the invisible danger in an epoch where vaccines did not exist, I had the privilege of working safely from home. This work is dedicated to those brave people. On a similar note, this work represents a milestone in my career. It is also dedicated to my former advisor Eduardo Cypriano, who patiently taught and guided me in the early stages of my career. I am incredibly grateful for his support.

To accomplish the results presented here, I used the R software. For more details, please visit https://www.r-project.org/

This work make used of data from the Hyper Suprime-Cam (HSC) collaboration. The Hyper Suprime-Cam Subaru Strategic Program (HSC-SSP) is led by the astronomical communities of Japan and Taiwan, and Princeton University.  The instrumentation and software were developed by the National Astronomical Observatory of Japan (NAOJ), the Kavli Institute for the Physics and Mathematics of the Universe (Kavli IPMU), the University of Tokyo, the High Energy Accelerator Research Organization (KEK), the Academia Sinica Institute for Astronomy and Astrophysics in Taiwan (ASIAA), and Princeton University.  The survey was made possible by funding contributed by the Ministry of Education, Culture, Sports, Science and Technology (MEXT), the Japan Society for the Promotion of Science (JSPS),  (Japan Science and Technology Agency (JST),  the Toray Science Foundation, NAOJ, Kavli IPMU, KEK, ASIAA,  and Princeton University. For more details, please visit https://hsc-release.mtk.nao.ac.jp/doc/index.php/acknowledging-hsc-2/

\section*{Data Availability}

All data used in this work is already publicity available. However, the processed data can be shared under request to the author.



\bibliographystyle{mnras}
\bibliography{monteiro-oliveira_library} 

\begin{thebibliography}{}
\makeatletter
\relax
\def\mn@urlcharsother{\let\do\@makeother \do\$\do\&\do\#\do\^\do\_\do\%\do\~}
\def\mn@doi{\begingroup\mn@urlcharsother \@ifnextchar [ {\mn@doi@}
  {\mn@doi@[]}}
\def\mn@doi@[#1]#2{\def\@tempa{#1}\ifx\@tempa\@empty \href
  {http://dx.doi.org/#2} {doi:#2}\else \href {http://dx.doi.org/#2} {#1}\fi
  \endgroup}
\def\mn@eprint#1#2{\mn@eprint@#1:#2::\@nil}
\def\mn@eprint@arXiv#1{\href {http://arxiv.org/abs/#1} {{\tt arXiv:#1}}}
\def\mn@eprint@dblp#1{\href {http://dblp.uni-trier.de/rec/bibtex/#1.xml}
  {dblp:#1}}
\def\mn@eprint@#1:#2:#3:#4\@nil{\def\@tempa {#1}\def\@tempb {#2}\def\@tempc
  {#3}\ifx \@tempc \@empty \let \@tempc \@tempb \let \@tempb \@tempa \fi \ifx
  \@tempb \@empty \def\@tempb {arXiv}\fi \@ifundefined
  {mn@eprint@\@tempb}{\@tempb:\@tempc}{\expandafter \expandafter \csname
  mn@eprint@\@tempb\endcsname \expandafter{\@tempc}}}

\bibitem[\protect\citeauthoryear{{Aihara} et~al.,}{{Aihara}
  et~al.}{2019}]{HSCSSP.PDR2}
{Aihara} H.,  et~al., 2019, \mn@doi [\pasj] {10.1093/pasj/psz103}, \href
  {https://ui.adsabs.harvard.edu/abs/2019PASJ...71..114A} {71, 114}

\bibitem[\protect\citeauthoryear{{Beers}, {Geller}  \& {Huchra}}{{Beers}
  et~al.}{1982}]{beers82}
{Beers} T.~C.,  {Geller} M.~J.,   {Huchra} J.~P.,  1982, \mn@doi [\apj]
  {10.1086/159958}, \href {http://adsabs.harvard.edu/abs/1982ApJ...257...23B}
  {257, 23}

\bibitem[\protect\citeauthoryear{{Bertin} \& {Arnouts}}{{Bertin} \&
  {Arnouts}}{1996}]{sextractor}
{Bertin} E.,  {Arnouts} S.,  1996, \aaps, \href
  {http://adsabs.harvard.edu/abs/1996A%26AS..117..393B} {117, 393}

\bibitem[\protect\citeauthoryear{{Bridle}, {Hobson}, {Lasenby}  \&
  {Saunders}}{{Bridle} et~al.}{1998}]{bridle98}
{Bridle} S.~L.,  {Hobson} M.~P.,  {Lasenby} A.~N.,   {Saunders} R.,  1998,
  \mn@doi [\mnras] {10.1046/j.1365-8711.1998.01877.x}, \href
  {http://adsabs.harvard.edu/abs/1998MNRAS.299..895B} {299, 895}

\bibitem[\protect\citeauthoryear{{Chadayammuri}, {ZuHone}, {Nulsen}, {Nagai},
  {Felix}, {Andrade-Santos}, {King}  \& {Russell}}{{Chadayammuri}
  et~al.}{2022}]{Chadayammuri21}
{Chadayammuri} U.,  {ZuHone} J.,  {Nulsen} P.,  {Nagai} D.,  {Felix} S.,
  {Andrade-Santos} F.,  {King} L.,   {Russell} H.,  2022, \mn@doi [\mnras]
  {10.1093/mnras/stab2629}, \href
  {https://ui.adsabs.harvard.edu/abs/2022MNRAS.509.1201C} {509, 1201}

\bibitem[\protect\citeauthoryear{{Cho}, {Jee}, {Smith}, {Finner}  \&
  {Lee}}{{Cho} et~al.}{2021}]{Cho21}
{Cho} H.,  {Jee} M.~J.,  {Smith} R.,  {Finner} K.,   {Lee} W.,  2021, arXiv
  e-prints, \href {https://ui.adsabs.harvard.edu/abs/2021arXiv210906879C} {p.
  arXiv:2109.06879}

\bibitem[\protect\citeauthoryear{{Chon}, {B{\"o}hringer}  \& {Zaroubi}}{{Chon}
  et~al.}{2015}]{chon15b}
{Chon} G.,  {B{\"o}hringer} H.,   {Zaroubi} S.,  2015, \mn@doi [\aap]
  {10.1051/0004-6361/201425591}, \href
  {https://ui.adsabs.harvard.edu/abs/2015A&A...575L..14C} {575, L14}

\bibitem[\protect\citeauthoryear{{Dawson}}{{Dawson}}{2013}]{dawson}
{Dawson} W.~A.,  2013, \mn@doi [\apj] {10.1088/0004-637X/772/2/131}, \href
  {http://adsabs.harvard.edu/abs/2013ApJ...772..131D} {772, 131}

\bibitem[\protect\citeauthoryear{{Dietrich}, {B{\"o}hnert}, {Lombardi},
  {Hilbert}  \& {Hartlap}}{{Dietrich} et~al.}{2012}]{dietrich12}
{Dietrich} J.~P.,  {B{\"o}hnert} A.,  {Lombardi} M.,  {Hilbert} S.,   {Hartlap}
  J.,  2012, \mn@doi [\mnras] {10.1111/j.1365-2966.2011.19995.x}, \href
  {http://adsabs.harvard.edu/abs/2012MNRAS.419.3547D} {419, 3547}

\bibitem[\protect\citeauthoryear{{Doubrawa}, {Machado}, {Lagan{\'a}}, {Lima
  Neto}, {Monteiro-Oliveira}  \& {Cypriano}}{{Doubrawa}
  et~al.}{2020}]{Doubrawa20}
{Doubrawa} L.,  {Machado} R.~E.~G.,  {Lagan{\'a}} T.~F.,  {Lima Neto} G.~B.,
  {Monteiro-Oliveira} R.,   {Cypriano} E.~S.,  2020, \mn@doi [\mnras]
  {10.1093/mnras/staa1051}, \href
  {https://ui.adsabs.harvard.edu/abs/2020MNRAS.495.2022D} {495, 2022}

\bibitem[\protect\citeauthoryear{{Duffy}, {Schaye}, {Kay}  \& {Dalla
  Vecchia}}{{Duffy} et~al.}{2008}]{duffy08}
{Duffy} A.~R.,  {Schaye} J.,  {Kay} S.~T.,   {Dalla Vecchia} C.,  2008, \mn@doi
  [\mnras] {10.1111/j.1745-3933.2008.00537.x}, \href
  {http://adsabs.harvard.edu/abs/2008MNRAS.390L..64D} {390, L64}

\bibitem[\protect\citeauthoryear{{Fischer}, {Br{\"u}ggen}, {Schmidt-Hoberg},
  {Dolag}, {Ragagnin}  \& {Robertson}}{{Fischer} et~al.}{2021}]{Fischer21}
{Fischer} M.~S.,  {Br{\"u}ggen} M.,  {Schmidt-Hoberg} K.,  {Dolag} K.,
  {Ragagnin} A.,   {Robertson} A.,  2021, \mn@doi [\mnras]
  {10.1093/mnras/stab3544}, \href
  {https://ui.adsabs.harvard.edu/abs/2021MNRAS.tmp.3203F} {}

\bibitem[\protect\citeauthoryear{{Ghirardini} et~al.,}{{Ghirardini}
  et~al.}{2021}]{Ghirardini21}
{Ghirardini} V.,  et~al., 2021, \mn@doi [\aap] {10.1051/0004-6361/202039554},
  \href {https://ui.adsabs.harvard.edu/abs/2021A&A...647A...4G} {647, A4}

\bibitem[\protect\citeauthoryear{{Ha}, {Ryu}  \& {Kang}}{{Ha}
  et~al.}{2018}]{Ha18}
{Ha} J.-H.,  {Ryu} D.,   {Kang} H.,  2018, \mn@doi [\apj]
  {10.3847/1538-4357/aab4a2}, \href
  {https://ui.adsabs.harvard.edu/abs/2018ApJ...857...26H} {857, 26}

\bibitem[\protect\citeauthoryear{{Harvey}, {Massey}, {Kitching}, {Taylor}  \&
  {Tittley}}{{Harvey} et~al.}{2015}]{harvey15}
{Harvey} D.,  {Massey} R.,  {Kitching} T.,  {Taylor} A.,   {Tittley} E.,  2015,
  \mn@doi [Science] {10.1126/science.1261381}, \href
  {http://adsabs.harvard.edu/abs/2015Sci...347.1462H} {347, 1462}

\bibitem[\protect\citeauthoryear{{Hern{\'a}ndez-Lang}
  et~al.,}{{Hern{\'a}ndez-Lang} et~al.}{2021}]{Hernandez-Lang22}
{Hern{\'a}ndez-Lang} D.,  et~al., 2021, arXiv e-prints, \href
  {https://ui.adsabs.harvard.edu/abs/2021arXiv211115443H} {p. arXiv:2111.15443}

\bibitem[\protect\citeauthoryear{{Hetterscheidt}, {Erben}, {Schneider},
  {Maoli}, {van Waerbeke}  \& {Mellier}}{{Hetterscheidt}
  et~al.}{2005}]{hetterscheidt05}
{Hetterscheidt} M.,  {Erben} T.,  {Schneider} P.,  {Maoli} R.,  {van Waerbeke}
  L.,   {Mellier} Y.,  2005, \mn@doi [\aap] {10.1051/0004-6361:20053339}, \href
  {http://adsabs.harvard.edu/abs/2005A%26A...442...43H} {442, 43}

\bibitem[\protect\citeauthoryear{{Kass} \& {Raftery}}{{Kass} \&
  {Raftery}}{1995}]{kass95}
{Kass} R.~E.,  {Raftery} A.~E.,  1995, Journal of the American Statistical
  Association, 90, 773

\bibitem[\protect\citeauthoryear{{Kelkar} et~al.,}{{Kelkar}
  et~al.}{2020}]{Kelkar20}
{Kelkar} K.,  et~al., 2020, \mn@doi [\mnras] {10.1093/mnras/staa1547}, \href
  {https://ui.adsabs.harvard.edu/abs/2020MNRAS.496..442K} {496, 442}

\bibitem[\protect\citeauthoryear{{Keshet}, {Raveh}  \& {Naor}}{{Keshet}
  et~al.}{2021}]{Keshet21}
{Keshet} U.,  {Raveh} I.,   {Naor} Y.,  2021, \mn@doi [\mnras]
  {10.1093/mnras/stab2808}, \href
  {https://ui.adsabs.harvard.edu/abs/2021MNRAS.508.3455K} {508, 3455}

\bibitem[\protect\citeauthoryear{{Knowles} et~al.,}{{Knowles}
  et~al.}{2021}]{Knowles22}
{Knowles} K.,  et~al., 2021, arXiv e-prints, \href
  {https://ui.adsabs.harvard.edu/abs/2021arXiv211105673K} {p. arXiv:2111.05673}

\bibitem[\protect\citeauthoryear{{Kuchner} et~al.,}{{Kuchner}
  et~al.}{2022}]{Kuchner22}
{Kuchner} U.,  et~al., 2022, \mn@doi [\mnras] {10.1093/mnras/stab3419}, \href
  {https://ui.adsabs.harvard.edu/abs/2022MNRAS.510..581K} {510, 581}

\bibitem[\protect\citeauthoryear{{Leauthaud} et~al.,}{{Leauthaud}
  et~al.}{2007}]{Leauthaud07}
{Leauthaud} A.,  et~al., 2007, \mn@doi [\apjs] {10.1086/516598}, \href
  {https://ui.adsabs.harvard.edu/abs/2007ApJS..172..219L} {172, 219}

\bibitem[\protect\citeauthoryear{{Liu} \& {Haiman}}{{Liu} \&
  {Haiman}}{2016}]{Liu16}
{Liu} J.,  {Haiman} Z.,  2016, \mn@doi [\prd] {10.1103/PhysRevD.94.043533},
  \href {http://adsabs.harvard.edu/abs/2016PhRvD..94d3533L} {94, 043533}

\bibitem[\protect\citeauthoryear{{Machado} \& {Lima Neto}}{{Machado} \& {Lima
  Neto}}{2013}]{Machado13}
{Machado} R.~E.~G.,  {Lima Neto} G.~B.,  2013, \mn@doi [\mnras]
  {10.1093/mnras/stt127}, \href
  {http://adsabs.harvard.edu/abs/2013MNRAS.430.3249M} {430, 3249}

\bibitem[\protect\citeauthoryear{{Machado}, {Monteiro-Oliveira}, {Lima Neto}
  \& {Cypriano}}{{Machado} et~al.}{2015}]{Machado+2015}
{Machado} R.~E.~G.,  {Monteiro-Oliveira} R.,  {Lima Neto} G.~B.,   {Cypriano}
  E.~S.,  2015, \mn@doi [\mnras] {10.1093/mnras/stv1162}, \href
  {http://adsabs.harvard.edu/abs/2015MNRAS.451.3309M} {451, 3309}

\bibitem[\protect\citeauthoryear{{Machado}, {Lagan{\'a}}, {Souza}, {Caproni},
  {Antas}  \& {Mello-Terencio}}{{Machado} et~al.}{2022}]{Machado22}
{Machado} R. E.~G.,  {Lagan{\'a}} T.~F.,  {Souza} G.~S.,  {Caproni} A.,
  {Antas} A. S.~R.,   {Mello-Terencio} E.~A.,  2022, arXiv e-prints, \href
  {https://ui.adsabs.harvard.edu/abs/2022arXiv220614127M} {p. arXiv:2206.14127}

\bibitem[\protect\citeauthoryear{{Marshall}, {Hobson}, {Gull}  \&
  {Bridle}}{{Marshall} et~al.}{2002}]{LensEnt2}
{Marshall} P.~J.,  {Hobson} M.~P.,  {Gull} S.~F.,   {Bridle} S.~L.,  2002,
  \mn@doi [\mnras] {10.1046/j.1365-8711.2002.05685.x}, \href
  {http://adsabs.harvard.edu/abs/2002MNRAS.335.1037M} {335, 1037}

\bibitem[\protect\citeauthoryear{{Martel}, {Robichaud}  \& {Barai}}{{Martel}
  et~al.}{2014}]{martel14}
{Martel} H.,  {Robichaud} F.,   {Barai} P.,  2014, \mn@doi [\apj]
  {10.1088/0004-637X/786/2/79}, \href
  {http://adsabs.harvard.edu/abs/2014ApJ...786...79M} {786, 79}

\bibitem[\protect\citeauthoryear{Martin, Quinn  \& Park}{Martin
  et~al.}{2011}]{MCMCpack}
Martin A.~D.,  Quinn K.~M.,   Park J.~H.,  2011, Journal of Statistical
  Software, 42, 22

\bibitem[\protect\citeauthoryear{{Medezinski}, {Broadhurst}, {Umetsu}, {Oguri},
  {Rephaeli}  \& {Ben{\'{\i}}tez}}{{Medezinski} et~al.}{2010}]{med10}
{Medezinski} E.,  {Broadhurst} T.,  {Umetsu} K.,  {Oguri} M.,  {Rephaeli} Y.,
  {Ben{\'{\i}}tez} N.,  2010, \mn@doi [\mnras]
  {10.1111/j.1365-2966.2010.16491.x}, \href
  {http://adsabs.harvard.edu/abs/2010MNRAS.405..257M} {405, 257}

\bibitem[\protect\citeauthoryear{{Medezinski} et~al.,}{{Medezinski}
  et~al.}{2018}]{Medezinski18}
{Medezinski} E.,  et~al., 2018, \mn@doi [\pasj] {10.1093/pasj/psy009}, \href
  {http://adsabs.harvard.edu/abs/2018PASJ...70...30M} {70, 30}

\bibitem[\protect\citeauthoryear{{Merloni} et~al.,}{{Merloni}
  et~al.}{2012}]{Merloni12}
{Merloni} A.,  et~al., 2012, arXiv e-prints, \href
  {https://ui.adsabs.harvard.edu/abs/2012arXiv1209.3114M} {p. arXiv:1209.3114}

\bibitem[\protect\citeauthoryear{{Molnar}, {Ueda}  \& {Umetsu}}{{Molnar}
  et~al.}{2020}]{Molnar20}
{Molnar} S.~M.,  {Ueda} S.,   {Umetsu} K.,  2020, \mn@doi [\apj]
  {10.3847/1538-4357/abac53}, \href
  {https://ui.adsabs.harvard.edu/abs/2020ApJ...900..151M} {900, 151}

\bibitem[\protect\citeauthoryear{{Monteiro-Oliveira}, {Cypriano}, {Machado},
  {Lima Neto}, {Ribeiro}, {Sodr{\'e}}  \& {Dupke}}{{Monteiro-Oliveira}
  et~al.}{2017a}]{Monteiro-Oliveira17a}
{Monteiro-Oliveira} R.,  {Cypriano} E.~S.,  {Machado} R.~E.~G.,  {Lima Neto}
  G.~B.,  {Ribeiro} A.~L.~B.,  {Sodr{\'e}} L.,   {Dupke} R.,  2017a, \mn@doi
  [\mnras] {10.1093/mnras/stw3238}, \href
  {http://adsabs.harvard.edu/abs/2017MNRAS.466.2614M} {466, 2614}

\bibitem[\protect\citeauthoryear{{Monteiro-Oliveira}, {Lima Neto}, {Cypriano},
  {Machado}, {Capelato}, {Lagan{\'a}}, {Durret}  \&
  {Bagchi}}{{Monteiro-Oliveira} et~al.}{2017b}]{Monteiro-Oliveira17b}
{Monteiro-Oliveira} R.,  {Lima Neto} G.~B.,  {Cypriano} E.~S.,  {Machado}
  R.~E.~G.,  {Capelato} H.~V.,  {Lagan{\'a}} T.~F.,  {Durret} F.,   {Bagchi}
  J.,  2017b, \mn@doi [\mnras] {10.1093/mnras/stx791}, \href
  {http://adsabs.harvard.edu/abs/2017MNRAS.468.4566M} {468, 4566}

\bibitem[\protect\citeauthoryear{{Monteiro-Oliveira}, {Cypriano}, {Vitorelli},
  {Ribeiro}, {Sodr{\'e}}, {Dupke}  \& {Mendes de Oliveira}}{{Monteiro-Oliveira}
  et~al.}{2018}]{Monteiro-Oliveira18}
{Monteiro-Oliveira} R.,  {Cypriano} E.~S.,  {Vitorelli} A.~Z.,  {Ribeiro}
  A.~L.~B.,  {Sodr{\'e}} L.,  {Dupke} R.,   {Mendes de Oliveira} C.,  2018,
  \mn@doi [\mnras] {10.1093/mnras/sty2349}, \href
  {http://adsabs.harvard.edu/abs/2018MNRAS.481.1097M} {481, 1097}

\bibitem[\protect\citeauthoryear{{Monteiro-Oliveira}, {Doubrawa}, {Machado},
  {Lima Neto}, {Castejon}  \& {Cypriano}}{{Monteiro-Oliveira}
  et~al.}{2020}]{Monteiro-Oliveira20}
{Monteiro-Oliveira} R.,  {Doubrawa} L.,  {Machado} R.~E.~G.,  {Lima Neto}
  G.~B.,  {Castejon} M.,   {Cypriano} E.~S.,  2020, \mn@doi [\mnras]
  {10.1093/mnras/staa1218}, \href
  {https://ui.adsabs.harvard.edu/abs/2020MNRAS.495.2007M} {495, 2007}

\bibitem[\protect\citeauthoryear{{Monteiro-Oliveira}, {Soja}, {Ribeiro},
  {Bagchi}, {Sankhyayan}, {Candido}  \& {Flores}}{{Monteiro-Oliveira}
  et~al.}{2021}]{Monteiro-Oliveira21}
{Monteiro-Oliveira} R.,  {Soja} A.~C.,  {Ribeiro} A.~L.~B.,  {Bagchi} J.,
  {Sankhyayan} S.,  {Candido} T.~O.,   {Flores} R.~R.,  2021, \mn@doi [\mnras]
  {10.1093/mnras/staa3575}, \href
  {https://ui.adsabs.harvard.edu/abs/2021MNRAS.501..756M} {501, 756}

\bibitem[\protect\citeauthoryear{{Monteiro-Oliveira}, {Morell}, {Sampaio},
  {Ribeiro}  \& {de Carvalho}}{{Monteiro-Oliveira}
  et~al.}{2022}]{Monteiro-Oliveira22}
{Monteiro-Oliveira} R.,  {Morell} D.~F.,  {Sampaio} V.~M.,  {Ribeiro} A.~L.~B.,
    {de Carvalho} R.~R.,  2022, \mn@doi [\mnras] {10.1093/mnras/stab3225},
  \href {https://ui.adsabs.harvard.edu/abs/2022MNRAS.509.3470M} {509, 3470}

\bibitem[\protect\citeauthoryear{{Moura}, {Machado}  \&
  {Monteiro-Oliveira}}{{Moura} et~al.}{2021}]{Moura21}
{Moura} M.~T.,  {Machado} R. E.~G.,   {Monteiro-Oliveira} R.,  2021, \mn@doi
  [\mnras] {10.1093/mnras/staa3399}, \href
  {https://ui.adsabs.harvard.edu/abs/2021MNRAS.500.1858M} {500, 1858}

\bibitem[\protect\citeauthoryear{{Navarro}, {Frenk}  \& {White}}{{Navarro}
  et~al.}{1996}]{nfw96}
{Navarro} J.~F.,  {Frenk} C.~S.,   {White} S.~D.~M.,  1996, \mn@doi [\apj]
  {10.1086/177173}, \href {http://adsabs.harvard.edu/abs/1996ApJ...462..563N}
  {462, 563}

\bibitem[\protect\citeauthoryear{{Nishizawa}, {Hsieh}, {Tanaka}  \&
  {Takata}}{{Nishizawa} et~al.}{2020}]{Nishizawa20}
{Nishizawa} A.~J.,  {Hsieh} B.-C.,  {Tanaka} M.,   {Takata} T.,  2020, arXiv
  e-prints, \href {https://ui.adsabs.harvard.edu/abs/2020arXiv200301511N} {p.
  arXiv:2003.01511}

\bibitem[\protect\citeauthoryear{{Nychka}, {Furrer}  \& {Sain}}{{Nychka}
  et~al.}{2014}]{fields}
{Nychka} D.,  {Furrer} R.,   {Sain} S.,  2014, Fields: Tools for spatial data.
\url {http://CRAN.R-project.org/package=fields}

\bibitem[\protect\citeauthoryear{{Pandge}, {Monteiro-Oliveira}, {Bagchi},
  {Simionescu}, {Limousin}  \& {Raychaudhury}}{{Pandge}
  et~al.}{2019}]{Pandge19}
{Pandge} M.~B.,  {Monteiro-Oliveira} R.,  {Bagchi} J.,  {Simionescu} A.,
  {Limousin} M.,   {Raychaudhury} S.,  2019, \mn@doi [\mnras]
  {10.1093/mnras/sty2937}, \href
  {https://ui.adsabs.harvard.edu/abs/2019MNRAS.482.5093P} {482, 5093}

\bibitem[\protect\citeauthoryear{{Plummer}, {Best}, {Cowles}  \&
  {Vines}}{{Plummer} et~al.}{2006}]{coda}
{Plummer} M.,  {Best} N.,  {Cowles} K.,   {Vines} K.,  2006, R News, 6, 7

\bibitem[\protect\citeauthoryear{{Predehl} et~al.,}{{Predehl}
  et~al.}{2021}]{eROSITA}
{Predehl} P.,  et~al., 2021, \mn@doi [\aap] {10.1051/0004-6361/202039313},
  \href {https://ui.adsabs.harvard.edu/abs/2021A&A...647A...1P} {647, A1}

\bibitem[\protect\citeauthoryear{{R Core Team}}{{R Core Team}}{2014}]{R}
{R Core Team} 2014, R: A Language and Environment for Statistical Computing.
R Foundation for Statistical Computing, Vienna, Austria, \url
  {http://www.R-project.org/}

\bibitem[\protect\citeauthoryear{{Sarazin}}{{Sarazin}}{2004}]{sarazin04}
{Sarazin} C.~L.,  2004, \mn@doi [Journal of Korean Astronomical Society]
  {10.5303/JKAS.2004.37.5.433}, \href
  {http://adsabs.harvard.edu/abs/2004JKAS...37..433S} {37, 433}

\bibitem[\protect\citeauthoryear{{Schirmer}}{{Schirmer}}{2004}]{Schirmer04}
{Schirmer} M.,  2004, PhD thesis, Rheinischen Friedrich-Wilhelms-Universit\"at
  Bonn

\bibitem[\protect\citeauthoryear{{Schneider}}{{Schneider}}{1996}]{schneider96}
{Schneider} P.,  1996, \mn@doi [\mnras] {10.1093/mnras/283.3.837}, \href
  {http://adsabs.harvard.edu/abs/1996MNRAS.283..837S} {283, 837}

\bibitem[\protect\citeauthoryear{{Soja}, {Sodr{\'e}}, {Monteiro-Oliveira},
  {Cypriano}  \& {Lima Neto}}{{Soja} et~al.}{2018}]{Soja18}
{Soja} A.~C.,  {Sodr{\'e}} L.,  {Monteiro-Oliveira} R.,  {Cypriano} E.~S.,
  {Lima Neto} G.~B.,  2018, \mn@doi [\mnras] {10.1093/mnras/sty638}, \href
  {http://adsabs.harvard.edu/abs/2018MNRAS.477.3279S} {477, 3279}

\bibitem[\protect\citeauthoryear{{Stott}, {Pimbblet}, {Edge}, {Smith}  \&
  {Wardlow}}{{Stott} et~al.}{2009}]{Stott09}
{Stott} J.~P.,  {Pimbblet} K.~A.,  {Edge} A.~C.,  {Smith} G.~P.,   {Wardlow}
  J.~L.,  2009, \mn@doi [\mnras] {10.1111/j.1365-2966.2009.14477.x}, \href
  {https://ui.adsabs.harvard.edu/abs/2009MNRAS.394.2098S} {394, 2098}

\bibitem[\protect\citeauthoryear{{Tam} et~al.,}{{Tam} et~al.}{2020}]{Tam20b}
{Tam} S.-I.,  et~al., 2020, \mn@doi [\mnras] {10.1093/mnras/staa1828}, \href
  {https://ui.adsabs.harvard.edu/abs/2020MNRAS.496.4032T} {496, 4032}

\bibitem[\protect\citeauthoryear{{Tanaka} et~al.,}{{Tanaka}
  et~al.}{2018}]{Tanaka18}
{Tanaka} M.,  et~al., 2018, \mn@doi [\pasj] {10.1093/pasj/psx077}, \href
  {https://ui.adsabs.harvard.edu/abs/2018PASJ...70S...9T} {70, S9}

\bibitem[\protect\citeauthoryear{{Torri}, {Meneghetti}, {Bartelmann},
  {Moscardini}, {Rasia}  \& {Tormen}}{{Torri} et~al.}{2004}]{Torri04}
{Torri} E.,  {Meneghetti} M.,  {Bartelmann} M.,  {Moscardini} L.,  {Rasia} E.,
   {Tormen} G.,  2004, \mn@doi [\mnras] {10.1111/j.1365-2966.2004.07508.x},
  \href {https://ui.adsabs.harvard.edu/abs/2004MNRAS.349..476T} {349, 476}

\bibitem[\protect\citeauthoryear{{Ueda}, {Ichinohe}, {Molnar}, {Umetsu}  \&
  {Kitayama}}{{Ueda} et~al.}{2020}]{Ueda20}
{Ueda} S.,  {Ichinohe} Y.,  {Molnar} S.~M.,  {Umetsu} K.,   {Kitayama} T.,
  2020, \mn@doi [\apj] {10.3847/1538-4357/ab7bdc}, \href
  {https://ui.adsabs.harvard.edu/abs/2020ApJ...892..100U} {892, 100}

\bibitem[\protect\citeauthoryear{{Ueda}, {Umetsu}, {Ng}, {Ichinohe}, {Kitayama}
   \& {Molnar}}{{Ueda} et~al.}{2021}]{Ueda21}
{Ueda} S.,  {Umetsu} K.,  {Ng} F.,  {Ichinohe} Y.,  {Kitayama} T.,   {Molnar}
  S.~M.,  2021, \mn@doi [\apj] {10.3847/1538-4357/ac1f16}, \href
  {https://ui.adsabs.harvard.edu/abs/2021ApJ...922...81U} {922, 81}

\bibitem[\protect\citeauthoryear{{Umetsu}}{{Umetsu}}{2020}]{Umetsu20}
{Umetsu} K.,  2020, \mn@doi [\aapr] {10.1007/s00159-020-00129-w}, \href
  {https://ui.adsabs.harvard.edu/abs/2020A&ARv..28....7U} {28, 7}

\bibitem[\protect\citeauthoryear{{Visvanathan} \& {Sandage}}{{Visvanathan} \&
  {Sandage}}{1977}]{Visvanathan77}
{Visvanathan} N.,  {Sandage} A.,  1977, \mn@doi [\apj] {10.1086/155464}, \href
  {https://ui.adsabs.harvard.edu/abs/1977ApJ...216..214V} {216, 214}

\bibitem[\protect\citeauthoryear{{Wei}, {Li}, {Kang}, {Liu}, {Fan}, {Yuan}  \&
  {Pan}}{{Wei} et~al.}{2018}]{Wei18}
{Wei} C.,  {Li} G.,  {Kang} X.,  {Liu} X.,  {Fan} Z.,  {Yuan} S.,   {Pan} C.,
  2018, \mn@doi [\mnras] {10.1093/mnras/sty1268}, \href
  {http://adsabs.harvard.edu/abs/2018MNRAS.478.2987W} {478, 2987}

\bibitem[\protect\citeauthoryear{{Wen} \& {Han}}{{Wen} \& {Han}}{2013}]{wen13}
{Wen} Z.~L.,  {Han} J.~L.,  2013, \mn@doi [\mnras] {10.1093/mnras/stt1581},
  \href {https://ui.adsabs.harvard.edu/abs/2013MNRAS.436..275W} {436, 275}

\bibitem[\protect\citeauthoryear{{Wittman}, {Golovich}  \& {Dawson}}{{Wittman}
  et~al.}{2018}]{Wittman17}
{Wittman} D.,  {Golovich} N.,   {Dawson} W.~A.,  2018, \mn@doi [\apj]
  {10.3847/1538-4357/aaee77}, \href
  {https://ui.adsabs.harvard.edu/abs/2018ApJ...869..104W} {869, 104}

\bibitem[\protect\citeauthoryear{{Zhang}, {Churazov}, {Dolag}, {Forman}  \&
  {Zhuravleva}}{{Zhang} et~al.}{2020}]{Zhang20}
{Zhang} C.,  {Churazov} E.,  {Dolag} K.,  {Forman} W.~R.,   {Zhuravleva} I.,
  2020, \mn@doi [\mnras] {10.1093/mnras/staa1013}, \href
  {https://ui.adsabs.harvard.edu/abs/2020MNRAS.494.4539Z} {494, 4539}

\bibitem[\protect\citeauthoryear{{van Weeren}, {de Gasperin}, {Akamatsu},
  {Br{\"u}ggen}, {Feretti}, {Kang}, {Stroe}  \& {Zandanel}}{{van Weeren}
  et~al.}{2019}]{vanWeeren19}
{van Weeren} R.~J.,  {de Gasperin} F.,  {Akamatsu} H.,  {Br{\"u}ggen} M.,
  {Feretti} L.,  {Kang} H.,  {Stroe} A.,   {Zandanel} F.,  2019, \mn@doi [\ssr]
  {10.1007/s11214-019-0584-z}, \href
  {https://ui.adsabs.harvard.edu/abs/2019SSRv..215...16V} {215, 16}

\makeatother
\end{thebibliography}








\bsp	
\label{lastpage}
\end{document}